\documentclass[1p, final]{elsarticle}

\usepackage{subfig}
\usepackage{amsmath,bm}
\usepackage{color}
\usepackage{enumerate}
\usepackage{amssymb}
\usepackage{epstopdf}
\usepackage{epsfig}
\usepackage[table,xcdraw]{xcolor}

\usepackage{lineno,hyperref}
\modulolinenumbers[5]

\journal{Journal of Computational Physics}

\begin{document}

\begin{frontmatter}

\title{Dissipation-Based Continuation Method for Multiphase Flow in Heterogeneous Porous Media}

\author{Jiamin Jiang}
\cortext[mycorrespondingauthor]{Corresponding author}
\ead{jiamin66@stanford.edu}

\author{Hamdi A.\ Tchelepi}

\address{Department of Energy Resources Engineering, Stanford University}
\address{367 Panama Street, Green Earth Sciences Building, Rm 067, Stanford, CA 94305, USA}

\date{April 23, 2018}

\begin{abstract}

In reservoir simulation, solution of the coupled systems of nonlinear algebraic equations that are associated with fully-implicit (backward Euler) discretization is challenging. Having a robust and efficient nonlinear solver is necessary in order for reservoir simulation to serve as the primary tool for managing the recovery processes of large-scale reservoirs. However, there are several outstanding challenges that are intimately connected to the highly nonlinear nature of the problem. Given a set of sources and sinks, the variation in the total velocity can span many orders of magnitude due to extreme contrasts in the permeability field in large-scale subsurface porous formations. Moreover, multiple and complex saturation fronts must be properly resolved throughout the three-dimensional reservoir model of interest. Add to that numerical simulation studies entail making field-scale predictions over many decades, and the challenge of developing robust and efficient nonlinear solvers across a very wide parameter space becomes clear. Here, we develop a continuation method based on the use of a dissipation operator. We focus on nonlinear two-phase flow and transport in heterogeneous formations in the presence of viscous, gravitational, and capillary forces. The homotopy is constructed by adding numerical dissipation to the coupled discrete conservation equations. A continuation parameter is introduced to control the amount of dissipation. Numerical evidence of multi-dimensional models and detailed analysis of single-cell problems are used to explain how the dissipation operator improves the nonlinear convergence of the coupled system of equations. An adaptive strategy to determine the dissipation coefficient is proposed. The dissipation level is computed locally for each cell interface.
We demonstrate the efficiency of the dissipation-based continuation (DBC) nonlinear solver using several examples, including 1D scalar transport and 2D heterogeneous problems with fully-coupled flow and transport. The DBC solver has better convergence properties compared with the standard damped-Newton solvers used in reservoir simulation. 

\end{abstract}

\end{frontmatter}

\section{Introduction}

Numerical reservoir simulation is an essential tool for the management of subsurface resources, including oil and gas recovery, groundwater remediation, and $\mathrm{CO_2}$ subsurface sequestration. The partial differential equations (PDEs) that govern multiphase fluid flow and transport in heterogeneous porous media are highly nonlinear. The transport equations in the presence of viscous and gravitational forces are characterized by non-convex and non-monotonic flux functions (Brenier and Jaffré 1991). These properties make the development of robust, efficient, and accurate discretization and solution schemes quite challenging (Peaceman 2000; Lee et al., 2015).

Several temporal discretization methods are available to solve the mass conservation equations that describe coupled flow and transport (Aziz and Settari 1979). The use of explicit temporal schemes poses severe restrictions on the timestep size because the Courant-Friedrichs-Lewy (CFL) numbers can vary by several orders of magnitude across the computational domain (Forsyth and Sammon 1986; Jenny et al. 2009). Therefore, implicit temporal schemes, such as the fully-implicit method (FIM) and sequential-implicit methods (SIM) are preferred in practice. The resulting nonlinear discrete system of conservation equations is usually cast in residual form and solved using a Newton-based method. For a target timestep, a sequence of nonlinear (Newton) iterations is performed until the solution (within a specified tolerance) of the nonlinear algebraic equations is achieved.
Each iteration involves construction of the full Jacobian matrix and solution of the corresponding linear system of equations. In practice, reservoir simulators often suffer from serious difficulties in obtaining converged numerical solutions for the preselected timestep size. This is because even though FIM is unconditionally stable, the nonlinear Newton-based solver will not be unconditionally convergent 
(Cai et al. 2002; Skogestad et al. 2013).
When convergence failures are encountered, the most commonly used remedy is to restart the nonlinear solver with a smaller timestep size (Younis et al. 2010). This procedure often leads to timestep sizes that are very conservative resulting in unnecessarily long computational time and wasted computations.

For large timestep sizes, or poor initial guesses, the standard Newton method may fail to converge. 
Obtaining a suitable initial guess for the Newton method is referred to as globalization. Damping, or safeguarding the Newton updates is a globalization technique for enlarging the convergence radius of the nonlinear solver 
(Deuflhard 2004; Knoll and Keyes 2004; Brune et al. 2015).
A number of heuristic strategies have been devised to choose the damping factors. In numerical modeling of flow and transport in porous media, a common idea across these methods is to apply a cellwise (i.e., local) damping of the saturation(s) to ensure that the maximum absolute change in saturation remains below a certain value (e.g., 0.2). Appleyard chopping 
is a widely used heuristic that prevents large saturation changes when a fluid phase that was immobile at the previous iteration becomes mobile, and vice versa (ECLIPSE 2008; Younis et al. 2010).

The convergence rate of the Newton method applied to the discrete transport problem can be improved significantly with physics-based damping. In the work of Jenny et al. (2009), the non-convexity of the flux (fractional-flow) function is identified as the major cause of nonlinear convergence difficulties for viscous-dominated flow in heterogeneous models. They used the inflection point of the analytical flux function to guide the Newton updating of saturation field. Their scheme was proved to be unconditionally convergent for immiscible nonlinear two-phase flow without buoyancy. Wang and Tchelepi (2013) extended the strategy of Jenny et al. (2009) to two-phase problems in the presence of both viscous and gravitational forces. They proposed a trust-region Newton solver, in which the analytical flux function is divided into saturation trust regions delineated by the inflection, unit-flux, and end points. The nonlinear updates are performed such that two successive iterations cannot cross any trust-region boundary. If a crossing is detected, the update is chopped such that the new saturation lies at the appropriate trust-region boundary. Li and Tchelepi (2015) illustrated that the trust regions should be determined based on the numerical flux function as opposed to the analytical one. This is because the analytical form of the flux function does not fully capture the complex nonlinearity of the discrete system, especially in the presence of counter-current flow due to buoyancy and capillarity. The complex inflection lines and non-differentiability of the numerical flux across the co-current/counter-current boundaries are the root causes for nonlinear convergence problems. These critical features change the curvature of the residual function abruptly, and they can lead to overshoots, oscillations, or divergence of the Newton iterations (Li and Tchelepi 2015). Recently, M{\o}yner (2016) extended the trust-region strategy to systems where the inflection points that delineate different trust regions need not be pre-specified. Instead, these values are estimated during the solution process by projecting the updates along the Newton path, and then applying a flux-search algorithm. The proposed strategy has been demonstrated to deal with complex problems, such as three-phase flow (M{\o}yner 2016).

Homotopy continuation, which has been widely adopted in numerical algebraic geometry (Sommese and Wampler 2005) and bifurcation analysis (Keller 1977), provides an alternative way to globalize the Newton method. For a given nonlinear algebraic system to be solved, a homotopy between the target (given) system and a new system that is easier to solve is constructed. The new system is deformed gradually into the original one through a path tracking algorithm. The homotopy process leads to the solution of the original system of equations. Younis et al. (2010) developed a Continuation-Newton (CN) method, in which the continuation parameter is tied to the timestep size. The CN method generates a sequence of iterates in the augmented space (original space plus the timestep) that evolves toward the target timestep size. A key component of the CN framework is the development of a quantitative measure of the proximity to the homotopy path. Following the solution path too closely results in poor computational efficiency; on the other hand, a loose tolerance may produce iterates that are too far from the solution path requiring large numbers of Newton correction steps. Wang (2012) demonstrates that the nonlinear performance of CN can be sensitive to the heuristic parameters used, including the step-length and the neighborhood tolerance around the solution path. 

Recently, Cogswell and Szulczewski (2017) applied a phase-field formulation for the incompressible, immiscible, two-phase transport problem with no buoyancy or capillarity. The system is augmented to include surface tension, and a semi-implicit temporal discretization is employed based on a convex energy splitting scheme. Using a homotopy continuation, the macroscopic surface tension is progressively decreased after each Newton iteration, and the phase-field solution evolves toward the original problem.

For computational fluid dynamics (CFD) applications, Brown and Zingg (2016) designed a dissipation-based continuation (DBC) algorithm to solve steady-state problems using an aerodynamic flow solver. Their homotopy is constructed by adding numerical dissipation to the discrete conservation equations with a continuation parameter that controls the magnitude of the dissipation. In this paper, we employ the DBC approach to solve coupled two-phase flow and transport in heterogeneous porous media. We follow the ideas proposed by Brown and Zingg (2016); however, this work has some notable differences. The DBC approach is studied for the first time for multiphase flow in natural porous media; the investigation includes detailed analysis of single-cell and one-dimensional (1D) problems to explain why the dissipation operator improves the nonlinear convergence of the coupled system of equations. We also present multi-dimensional solutions using an adaptive strategy to determine the dissipation coefficient. 
%

\section{Immiscible multiphase flow and transport in porous media}

We consider compressible and immiscible flow and transport with $n_p$ fluid phases. The mass conservation equation for phase $\alpha$ is
\begin{equation} 
\frac{\partial }{\partial t }\left ( \phi \rho_{\alpha} S_{\alpha} \right ) + \nabla \cdot \left (\rho_{\alpha} \mathbf{u_{\alpha}} \right ) = \rho_{\alpha} q_{\alpha}
\label{eq:mass_con}
\end{equation}
where $\alpha \in \left \{ 1,...,n_p \right \}$. $\phi$ is the rock porosity. $\rho_\alpha$ and $\mathbf{u_\alpha}$ are the density and velocity of each phase, respectively. $q_{\alpha}$ is the well flow rate (source and sink terms). $S_\alpha$ is phase saturation, with the constraint that the sum of saturations is equal to one 
\begin{equation} 
\sum_{\alpha} S_{\alpha} = 1 
\label{eq:s_con}
\end{equation}
The phase velocities can be expressed using the multiphase extension of Darcy's law
\begin{equation}
\mathbf{u_\alpha} = - k \lambda_{\alpha} \left(\nabla p_{\alpha} + \rho_\alpha g \nabla h\right)
\label{eq:phase_v}
\end{equation}
where $k$ is the scalar rock permeability. $p_{\alpha}$ is phase pressure. $g$ is gravitational acceleration, and $h$ is height. The phase mobility $\lambda_{\alpha}(S_1,...,S_{n_p}) = k_{r\alpha}(S_1,...,S_{n_p})/\mu_\alpha$. $k_{r\alpha}$ and $\mu_\alpha$ are the relative permeability and viscosity, respectively. The capillary pressure constraint relates the phase pressures $p_{\alpha} \left ( \alpha \in \left \{ 1,...,n_p \right \} \right )$ to the reference pressure $p_{\alpha_0}$ as follows
\begin{equation}
P_{c, \alpha }(S_{\alpha}) = p_{\alpha_0} - p_{\alpha}
\end{equation}

\section{Fully-implicit finite-volume discretization}

The coupled multiphase problem in Eq.\ (\ref{eq:mass_con}) is highly nonlinear, and it can be challenging to solve the system for heterogeneous porous media. Let $\Omega \subset \mathbb{R}^{n_d}$ be the domain in dimension $n_d$. A standard finite-volume scheme is utilized as the spatial discretization for the mass conservation equations. The simulation grid represents a partition of $\Omega$ into a set of non-overlapping control volumes. The method of choice for the time discretization is often the first-order backward Euler scheme. Then, the fully-implicit discretization of a cell can be written as
\begin{equation}
R_A + R_F = R_W
\label{eq:resi}
\end{equation}
where the accumulation, flux and well parts are
\begin{equation}
R_A = \frac{\left | \Omega_i \right |}{\Delta t} \left ( \left ( \phi_i \rho_{\alpha,i}S_{\alpha,i} \right )^{n+1} - \left ( \phi_i \rho_{\alpha,i}S_{\alpha,i} \right )^{n} \right )
\end{equation}
\begin{equation}
R_F = \sum_{j\in adj(i)}\rho_{\alpha,ij}^{n+1}F_{\alpha,{ij}}^{n+1} \left ( \Delta p_{\alpha,{ij}},S_i,S_j \right )
\end{equation}
\begin{equation}
R_W = \rho_{\alpha,i}^{n+1}Q_{\alpha,i}^{n+1} \left ( p_{\alpha,i},S_i \right ) 
\end{equation}
where $i \in \left \{ 1,...,N \right \}$, and the shorthand notation $S_i = \left \{ S_{k,i} \right \}_{k \in \left \{ 1,...,n_p \right \}}$ refers to the saturations in cell $i$. $\Delta t$ is the timestep size. $\left | \Omega_i \right |$ is the volume of cell $i$. $adj(i)$ denotes the set of adjacent cells, such that cell $i$ shares an interface $\Gamma_{ij}$ with cell $j$. $\Delta p_{\alpha,{ij}}=p_{\alpha,i}-p_{\alpha,j}$ is the difference in the phase pressure between $i$ and $j$. For two-phase flow, the water saturation is chosen as a primary variable; for three-phase flow, the water and gas saturations are chosen as primary variables (Kwok and Tchelepi 2008).

In the finite-volume method, the discrete flux $F_{\alpha,ij}$ is constructed as the approximation of the velocity integrated over the interface $(ij)$ between two cells
\begin{equation}
F_{\alpha,ij} = T_{ij} \lambda_{\alpha,ij} \left ( \Delta p_{\alpha,{ij}} + \overline{\rho_\alpha} g \Delta h_{ij} \right )
\label{eq:dis_p_f}
\end{equation}
where $\overline{\rho_\alpha}$ is the arithmetic average for the phase densities of the two cells. The phase mobility $\lambda_{\alpha,ij}$ in the numerical flux is evaluated using the first-order upstream weighting scheme. 
The computation of the transmissibility $T_{ij}$ is based on the absolute permeability and mesh geometry in a stencil involving multiple cells around interface $(ij)$. We consider a two-point flux approximation (TPFA) of the transmissibility involving only cells $i$ and $j$. Combining two half-transmissibilities in a half of the harmonic average, we obtain
\begin{equation}
T_{ij} = \frac{T_i T_j}{T_i + T_j}
\end{equation}
The two-point half-transmissibility for a general unstructured mesh is evaluated by imposing flux and pressure continuity at the center of the interface. For a structured Cartesian grid, $T_{i}$ is defined as
\begin{equation}
T_{i} = \frac{k_i A_{ij}}{d_i}
\end{equation}
$A_{ij}$ is the area of interface $(ij)$ and $k_i$ denotes the absolute permeability in cell $i$. $d_i$ denotes the distance from the center of cell $i$ to interface $(ij)$.

\subsection{Newton method}

At each timestep of an implicit simulation, given the current state 
$\textit{\textbf{U}}^n$, and a fixed timestep size $\Delta t$, we seek to obtain the new state $\textit{\textbf{U}}^{n+1}$, by solving the corresponding nonlinear residual system using the Newton method
\begin{equation} 
\mathcal{R}(\textit{\textbf{U}}^{n+1}) = \textbf{0}
\end{equation}
The Newton method generates a sequence of iterates, $\textit{\textbf{U}}^{\nu}$, $\nu=0,1,...$, that would - ideally - converge to the correct solution. Specifically, successive linearization and updating are performed until convergence of the system of nonlinear residual equations is reached; that is
\begin{equation} 
\mathcal{J}(\textit{\textbf{U}}^{\nu }) \Delta \textit{\textbf{U}}^{\nu + 1} = - \mathcal{R}(\textit{\textbf{U}}^{\nu })
\end{equation}
where
\begin{equation} 
\Delta \textit{\textbf{U}}^{\nu + 1} = \textit{\textbf{U}}^{\nu +1} - \textit{\textbf{U}}^{\nu }
\end{equation}
and $\mathcal{J}(\textit{\textbf{U}}^{\nu }) = \left. \frac{\partial \mathcal{R}}{\partial \textit{\textbf{U}}} \right|_{\textit{\textbf{U}}^\nu }$ denotes the Jacobian matrix of $\mathcal{R}$ with respect to $\textit{\textbf{U}}^{\nu }$.

Usually, the initial guess to the Newton iteration of a new timestep is the solution of the previous timestep. For small timestep sizes, this is a reasonable starting point. For large timesteps, however, this may not be the case, and the Newton process may converge too slowly, or diverge (Younis et al. 2010).

\section{Upwinding scheme for immiscible two-phase transport}

Here, we consider the immiscible two-phase case (oil and water). We assume that the rock and the fluids are incompressible, then the total-velocity is a constant in one dimension, and the reduced problem can be written as a scalar hyperbolic conservation equation using the fractional-flow formulation (Chen et al. 2006). The elliptic pressure equation is obtained by summing the governing equations of the phases
\begin{equation} 
\nabla \cdot \mathbf{u}_T = q_T
\end{equation}
where $\mathbf{u}_T$ is the total velocity
\begin{equation} 
\mathbf{u}_T = \mathbf{u}_o+\mathbf{u}_w = -k \lambda_T \nabla p_w - k (\lambda_o \rho_o + \lambda_w \rho_w) g \nabla h -k \lambda_o \nabla P_c 
\end{equation}
where $P_c$ is the capillary pressure, which relates the pressures of the two phases. It is a highly nonlinear function of saturation, often expressed as $P_c(S)$
\begin{equation} 
P_c(S) = p_o - p_w
\end{equation}
The total mobility is defined as $\lambda_T = \lambda_o + \lambda_w$. From here on we let $S \equiv S_w$, and the water transport equation is written as
\begin{equation} 
\phi \frac{\partial S }{\partial t } + \nabla \cdot \left ( \frac{\lambda_w}{\lambda_T}\mathbf{u}_T -k \frac{\lambda_w \lambda_o}{\lambda_T} \left ( \rho_w - \rho_o \right )g \nabla h + k \frac{\lambda_w \lambda_o}{\lambda_T} \nabla P_c \right ) = q_w 
\label{Eq:scalar}
\end{equation}
Eq.\ (\ref{Eq:scalar}) is a degenerate parabolic equation in the presence of capillary pressure and a nonlinear hyperbolic equation when capillarity is neglected.

The dimensionless flux function (fractional flow) can be written as
\begin{equation} 
f_w = \frac{u_w}{u_T} = \frac{\lambda_w}{\lambda_T} - \frac{\lambda_w k_{ro}}{\lambda_T} \frac{C_g}{u_T} + \frac{\lambda_w k_{ro}}{\lambda_T} \frac{k \nabla P_c}{\mu_o u_T}
\end{equation}
where the three terms on the right-hand side represent the fluxes due to the viscous, gravitational, and capillary forces, respectively. The dimensionless gravity number $C_g$ is 
\begin{equation}
C_g = \frac{kg(\rho_w-\rho_o)}{\mu_o}\nabla h
\end{equation}
A dimensionless Peclet number ($P_e$) is defined as the ratio of the viscous and capillary forces
\begin{equation} 
P_e = \frac{u_T \mu_o L}{k \bar{P_c}}
\end{equation}
where $L$ is a characteristic length scale, and $\bar{P_c}$ is a characteristic capillary pressure. Therefore, the capillary flux component of $f_w$ is rewritten as
\begin{equation} 
f_c = \frac{\lambda_w k_{ro}}{\lambda_T} \frac{\nabla P_c}{\left ( \bar{P_c}/L \right ) P_e} = \left ( \frac{M k_{rw} k_{ro}}{M k_{rw} + k_{ro}} \frac{\textrm{d} P_c}{\textrm{d} S} \right ) \frac{1}{\left ( \bar{P_c}/L \right ) P_e} \nabla S 
\label{Eq:f_c_11}
\end{equation}
where $M$ is the viscosity ratio, $\mu_o/\mu_w$. In the absence of capillarity, the water phase velocity becomes
\begin{equation}
\label{Eq:uw}
u_w = \frac{Mk_{rw}}{k_{ro}+Mk_{rw}} u_T - \frac{Mk_{ro}k_{rw}}{k_{ro}+Mk_{rw}} C_g 
\end{equation}

\begin{figure}[!htb]
\centering
\subfloat[Viscous flux]{
\includegraphics[scale=0.35]{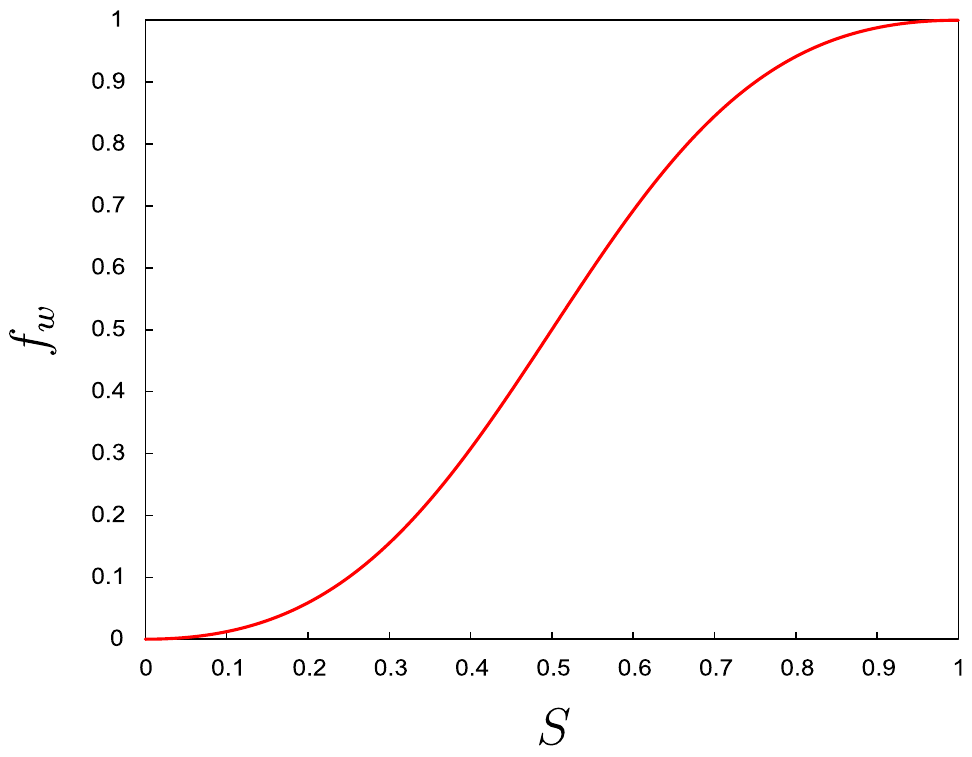}}
\subfloat[Viscous-gravitational flux]{
\includegraphics[scale=0.35]{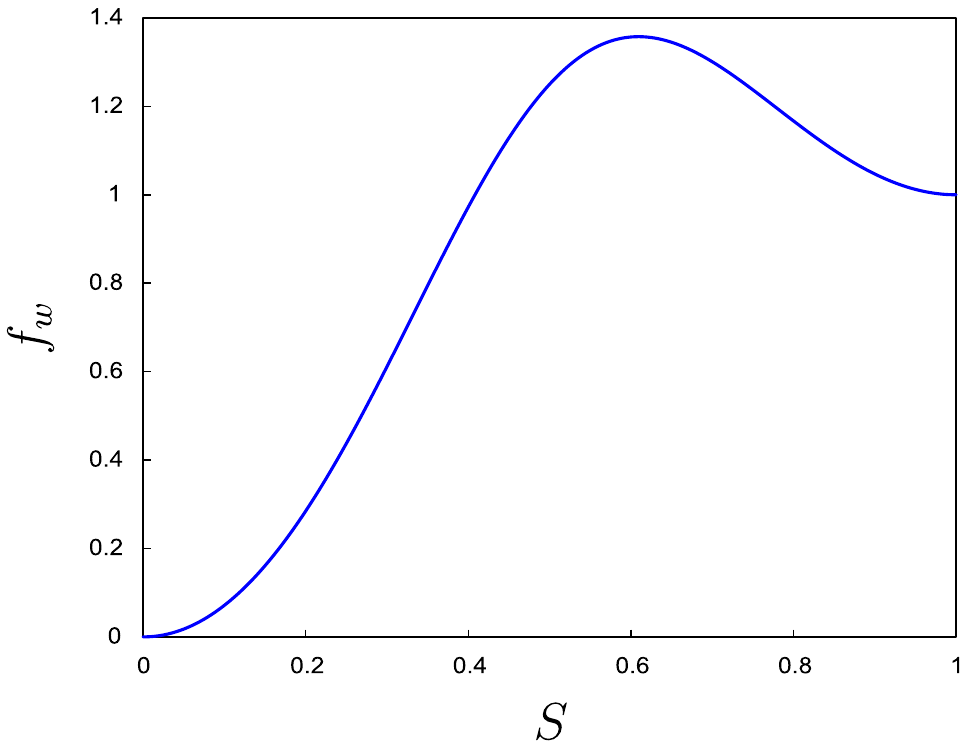}}
\caption{Fractional flow curves}
\label{fig:VG_flux}
\end{figure}

The behavior of the flux function depends on the magnitude of $u_T$ relative to $C_g$. For two-phase flow with viscous forces only, $f_w$ is S-shaped and has an inflection point. More complex dynamics develop when gravitational forces are dominant, in which case $f_w$ is usually a bell-shaped curve with two inflection points on either side of the sonic point. \textbf{Fig. \ref{fig:VG_flux}} demonstrates the fractional flow curves for viscous and viscous-gravitational forces with quadratic relative permeability functions.

The 1D scalar transport problem is discretized using the finite-volume formulation as
\begin{equation} 
\left ( S_{i}^{n+1} - S_{i}^{n} \right ) + \frac{\Delta t}{\phi \Delta x}\left ( F_{i,i+1}^{n+1} - F_{i-1,i}^{n+1} \right ) = 0
\label{Eq:1D_scheme}
\end{equation}
where $\Delta t$ is timestep and $\Delta x$ is cell size. Here, we define $S_i$ as the upstream state with respect to the total velocity $u_T$, and $S_j$ as the downstream state. For example, the upstream for $F_{ij}$ is cell $i$ if the direction of total velocity is from cell $i$ to $j$ ($u_T>0$). In this work, we always take the total velocity to be positive.

It has been shown that nonlinear properties of the numerical flux could have a large impact on the convergence behavior of the Newton-based iterative process (Wang and Tchelepi 2013). We review the Phase-Potential Upwinding (PPU) scheme, which is a popular numerical flux in reservoir simulation practice. In PPU, the upstream direction of a fluid phase is determined according to the gradient of the phase potential across the interface between two computational cells. The relative permeabilities at the cell interface ($ij$) in Eq.\ (\ref{Eq:uw}) are evaluated as 
\begin{equation} 
\label{eq:PPU}
k_{r\alpha,ij} = \left\{ {\begin{array}{*{20}c}
k_{r\alpha}(S_{i}),   & \Delta p_{ij} + \overline{\rho_\alpha} g \Delta h_{ij} > 0 \\ 
k_{r\alpha}(S_{j}), &  \mathrm{otherwise}
\end{array}} \right.  \ \ \alpha=w,o.
\end{equation}
From Eq.\ (\ref{eq:PPU}), we can rewrite the upstream conditions in terms of $u_T$ as
\begin{equation*} 
k_{rw,ij} = \left\{ {\begin{array}{*{20}c}
k_{rw}(S_{i}),   &   u_T - C_g k_{ro} > 0 \\ 
k_{rw}(S_{j}), &  \mathrm{otherwise}
\end{array}} \right.
\end{equation*}
and
\begin{equation} 
\label{Eq:up_c}
k_{ro,ij} = \left\{ {\begin{array}{*{20}c}
k_{ro}(S_{i}),   &   u_T + C_g M k_{rw} > 0 \\ 
k_{ro}(S_{j}), &  \mathrm{otherwise}
\end{array}} \right.
\end{equation}
Eq.\ (\ref{Eq:up_c}) does not explicitly define the upstream direction of $k_{r\alpha}$. Brenier and Jaffre (1991) used the following definition:
\begin{equation} 
k_{rw,ij} = \left\{ {\begin{array}{*{20}c}
k_{rw}(S_{i}),   &  \theta_w > 0 \\ 
k_{rw}(S_{j}), &  \mathrm{otherwise}
\end{array}} \right.
\end{equation}
where
\begin{equation} 
\label{Eq:theta_w}
\theta_w =  u_T - C_g k_{ro}(S_i) 
\end{equation}
and
\begin{equation} 
k_{ro,ij} = \left\{ {\begin{array}{*{20}c}
k_{ro}(S_{i}),   &  \theta_o > 0 \\ 
k_{ro}(S_{j}), &  \mathrm{otherwise}
\end{array}} \right.
\end{equation}
where
\begin{equation} 
\label{Eq:theta_o}
\theta_o = u_T + C_g M k_{rw}(S_i)
\end{equation}
In the transport iterations, because the total-velocity field is fixed, flow reversal from co-current to counter-current at a cell interface depends only on the saturation change in the upwind cell. From Eqs.\ (\ref{Eq:theta_w}) and (\ref{Eq:theta_o}), if $C_g > 0$ (updip), $\theta_o$ is always positive. Then, it follows that
\begin{equation} 
\label{Eq:kr_up}
\left\{ {\begin{array}{*{20}c}
k_{rw,ij} = k_{rw}(S_{i}) \ \mathrm{and} \ k_{ro,ij} = k_{ro}(S_{i}), & 0\leq \theta_w \leq \theta_o \\ 
k_{rw,ij} = k_{rw}(S_{j}) \ \mathrm{and} \ k_{ro,ij} = k_{ro}(S_{i}), & \theta_w \leq 0 \leq \theta_o 
\end{array}} \right.
\end{equation}
If $C_g < 0$ (downdip), now $\theta_w$ is always positive, and thus
\begin{equation} 
\label{Eq:kr_up_2}
\left\{ {\begin{array}{*{20}c}
k_{rw,ij} = k_{rw}(S_{i}) \ \mathrm{and} \ k_{ro,ij} = k_{ro}(S_{i}), & 0 \leq \theta_o \leq \theta_w \\ 
k_{rw,ij} = k_{rw}(S_{i}) \ \mathrm{and} \ k_{ro,ij} = k_{ro}(S_{j}), & \theta_o \leq 0 \leq \theta_w
\end{array}} \right.
\end{equation}
The PPU scheme is used to evaluate the combined viscous-gravitational term.\
The saturation-dependent coefficient in Eq.\ (\ref{Eq:f_c_11})
is computed using the arithmetic average of the two cells.

\section{Homotopy continuation method}

We consider the globalization technique in the class of homotopy continuation methods (Watson 1990; Allgower and Georg 1993). The objective is to solve $\mathcal{R}(\textit{\textbf{U}}) = \textbf{0}$. We introduce a continuation parameter $\kappa \in \left [ 0, 1 \right ]$ and a modified residual $\mathcal{H}(\textit{\textbf{U}},\kappa )$, such that $\mathcal{H}(\textit{\textbf{U}},0 ) = \mathcal{R}(\textit{\textbf{U}})$. The modified residual $\mathcal{H}$ is called a homotopy mapping. Moreover, assume that $\mathcal{H}(\textit{\textbf{U}},1 ) = \textbf{0}$ is significantly easier to solve than the target problem $\mathcal{H}(\textit{\textbf{U}},0 ) = \textbf{0}$. Geometrically representing the homotopy as a curve 
existing in the same real space as the flow equations, the homotopy continuation method can be developed by discretizing in $\kappa$ to form a sequence of nonlinear systems $\mathcal{H}(\textit{\textbf{U}}, \kappa_\eta ) = \textbf{0}$ (Brown and Zingg 2016). The curve is traced by progressively decreasing $\kappa$ from 1 to 0 to globalize the flow equations. In this way, the homotopy algorithm will not have an impact on the accuracy of the final solutions.

\begin{figure}[!htb]
\centering
\includegraphics[scale=0.7]{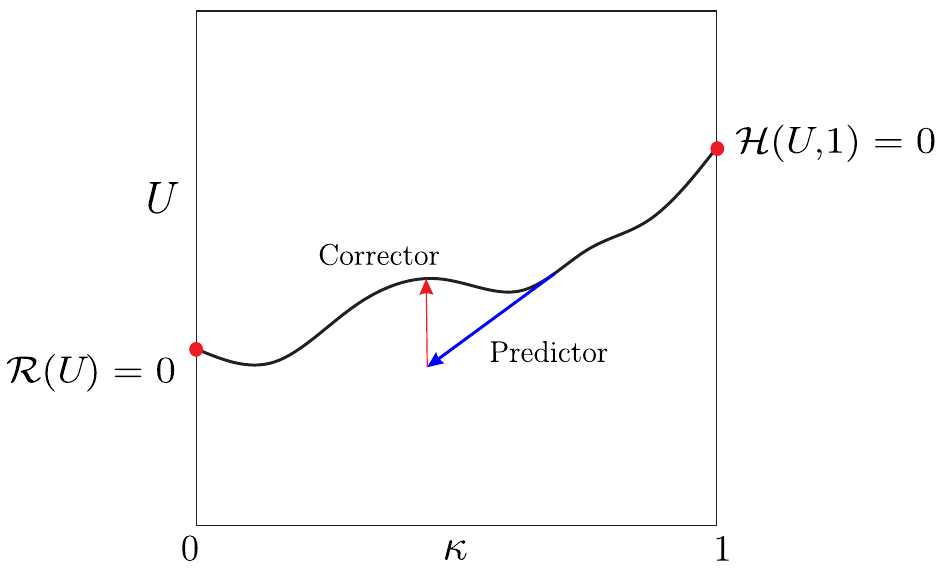}
\caption{Predictor-corrector technique for the problem with a single state unknown.}
\label{fig:homo}
\end{figure}

The solution path can be traced out using a predictor-corrector (PC) technique.
\textbf{Fig. \ref{fig:homo}} demonstrates the PC process for the nonlinear problem $\mathcal{H}(U,\kappa ) = 0$ with a single state unknown. The solution path is represented in the two dimensions, $U$ and $\kappa$. In a typical reservoir simulation problem, there may be millions of unknowns, and each would be represented by a dimension. The predictor and corrector steps are applied repeatedly until the solution is reached. These steps are:
1. Corrector: solve the nonlinear sub-problem at $\kappa_\eta$, to maintain the solution close enough to the homotopy path.
2. Predictor: obtain a suitable starting guess for the $(\eta+1)$-th sub-problem using the estimated solution from the $\eta$-th sub-problem, a step direction, and a step-length to travel in that direction. The choice of predictor is very important for both speed and robustness. Step-length adaptation algorithms are used to automatically adjust the step-length during traversing. The simplest method is to update $\kappa$, and use the solution to $\mathcal{H}(\textbf{\textit{U}},\kappa_\eta) = \textbf{0}$ as the initial guess at $\kappa_{\eta+1}$.

Younis et al. (2010) developed a Continuation-Newton (CN) method, in which the continuation parameter is tied to the timestep size. The CN method employs a PC strategy to follow the solution path in the augmented space toward the target timestep. It should be noted that the PC algorithm can be sometimes wasteful. This is because in order to ensure that the algorithm is convergent, the parameters are chosen such that the corrector phase can be solved, which can lead to `over-solving' along the way. 
Much of the work done in the corrector phase leads to only marginal improvement to the quality of the predictor update (Brown and Zingg 2017). 

Here, we apply a method that updates $\kappa$ and $\textit{\textbf{U}}$ simultaneously at each iteration. We view $\kappa$ as an independent variable, i.e., $\kappa$ plays the same role as the primary variables $\textit{\textbf{U}}$. We introduce a 
regularization function $\Phi(\kappa)$ that satisfies $\Phi(\kappa) \geq 0$, and $\Phi(\kappa) = 0$, if and only if $\kappa = 0$. Then, an augmented residual system is obtained as
\begin{equation} 
\mathcal{F}(\textit{\textbf{U}},\kappa) = \binom{\mathcal{H}(\textit{\textbf{U}},\kappa)}{\Phi(\kappa)}
\end{equation}
with the augmented updates $\Delta \textit{\textbf{Z}} = \left ( \Delta \textit{\textbf{U}} , \Delta \kappa \right )$ we seek to find at each iteration by the Newton linearization
\begin{equation} 
\renewcommand*{\arraystretch}{2.5}
\begin{bmatrix}
\dfrac{\partial \mathcal{H}}{\partial \textit{\textbf{U}}} & \dfrac{\partial \mathcal{H}}{\partial \kappa} \\ 
& \dfrac{\partial \Phi}{\partial \kappa}
\end{bmatrix}^{\nu }
\begin{bmatrix}
\Delta \textit{\textbf{U}} \\ 
\Delta \kappa
\end{bmatrix}^{\nu + 1}
= -\begin{bmatrix}
\mathcal{H}(\textit{\textbf{U}},\kappa) \\ 
\Phi(\kappa)
\end{bmatrix}^{\nu }
\label{Eq:aug_sys}
\end{equation}
There is no need to solve the linear system 
as a whole. The primary variables $\textit{\textbf{U}}$ can be directly updated from 
\begin{equation} 
\frac{\partial \mathcal{H}}{\partial \textit{\textbf{U}}} \Delta \textit{\textbf{U}}^{\nu + 1} = -\mathcal{H}(\textit{\textbf{U}}^{\nu },\kappa^{\nu }) - \frac{\partial \mathcal{H}}{\partial \kappa} \Delta \kappa^{\nu + 1}
\end{equation}
and the homotopy step size $\Delta \kappa$ is obtained through
\begin{equation} 
\frac{\partial \Phi}{\partial \kappa} \Delta \kappa^{\nu + 1} = -\Phi(\kappa^{\nu })
\label{Eq:kappa_up}
\end{equation}
In this procedure, $\Phi(\kappa)$ can be viewed as an updater for $\kappa$. From Eq.\ (\ref{Eq:kappa_up}) we can see that with a given updater, the update schedule of $\kappa$ is fixed for different flow problems. 

In this work, we implement a dissipation operator to construct the homotopy system. As will be illustrated in the result sections, the dissipation operator is quite effective in resolving an important type of nonlinear convergence difficulty. Here we do not insist on following the solution path closely. The continuation method only acts as a globalization stage to obtain better initial guesses for the Newton process. In our experience, applying the augmented system (\ref{Eq:aug_sys}) for a few iterations in the globalization stage is already very effective. The nonlinear performance of the dissipation-based continuation method is not sensitive to the update schedule of the continuation parameter. In other words, solving the discrete flow equations with addition of numerical dissipation is the key to achieving much improved nonlinear convergence.

For general homotopy systems, the evolution of the continuation parameter affects the efficiency and robustness of continuation methods. An efficient and systematic method to reduce the continuation parameter will be presented in a forthcoming paper.

\section{Dissipation operator}

For convection dominated flows, 
it is important to introduce some form of dissipation to stabilize the numerical solutions, either through the discretization scheme itself or as an additional term, known as artificial dissipation (Pulliam 1986). In this work, we employ a dissipation operator to construct an efficient homotopy continuation method. Consider the dissipation-based homotopy of the form (Brown and Zingg 2016)
\begin{equation} 
\mathcal{H}(\textit{\textbf{U}},\kappa ) = \mathcal{R}(\textit{\textbf{U}}) + \kappa \mathcal{D}(\textit{\textbf{U}})
\end{equation}
where $\mathcal{D}(\textit{\textbf{U}})$ is an
artificial dissipation term. For a simple 1D scalar transport case, the numerical flux in Eq.\ (\ref{Eq:1D_scheme}) can be expressed as the sum of a viscous-gravitational flux $F_{vg}$ and a dissipation flux $F_{d}$
\begin{equation} 
F_{ij} = F_{vg,ij} + F_{d,ij}
\end{equation}
where $F_{vg}$ is computed using the PPU scheme, and $F_{d}$ is given as
\begin{equation} 
F_{d,ij} = \varepsilon \frac{S_i - S_{j}}{ \Delta x}
\end{equation}

The previous implementation (Brown and Zingg 2016) of the dissipation operator is based on a finite-difference discretization. Here, we use a scalar dissipation, which is straightforward to implement in a finite-volume scheme, namely 
\begin{equation} 
F_{d,ij} = \kappa \, \beta \left ( S_i - S_{j} \right )
\label{Eq:d_scalar}
\end{equation}
where $\beta$ is a tunable coefficient to control the dissipation level. Note that in Eq. (\ref{Eq:d_scalar}) the cell size $\Delta x$ is incorporated into $\beta$, to introduce a better derivation. For sufficiently large values of $\beta$, the homotopy system is expected to be much easier to solve than the target system. In practice, $\beta$ can be computed in an adaptive way to determine the `optimal' level of dissipation before each timestep.

\section{1D examples: scalar transport problem}

We consider a 1D scalar transport with buoyancy. We assume that the total velocity and the gravity number are constant: $u_T = 0.01 $ and $C_g = 0.03 $. The viscosity ratio $M = 5 $, and quadratic relative-permeability functions are used: $k_{rw} = S^2$, $k_{ro} = (1-S)^2$. Flow is through a domain with unit length $x\in \left [ 0,1 \right ]$ and $N = 500 $ grid-blocks. The injection saturation is unity ($S_{inj} = 1 $) at the bottom (left) boundary and fluid is produced from the top (right) boundary. The timestep size is 5. 
The corresponding CFL number for this case is about 69. The initial condition is taken to be the initial guess for the Newton iteration. The damping strategy is employed to stabilize the Newton updates: the maximum absolute change in saturation remains as $\Delta S = 0.2 $. The following updater $\Phi(\kappa)$ for $\kappa$ is used
\begin{equation} 
\Phi(\kappa) = e^{\kappa}-1
\label{Eq:kappa_rule}
\end{equation}
Note that the chopping $\left ( \Delta \kappa \right )_{max} = 0.2 $ is also applied to gradually reduce $\kappa$.

\begin{figure}[!htb]
\centering
\subfloat[Initial condition $S=0 $]{
\includegraphics[scale=0.7]{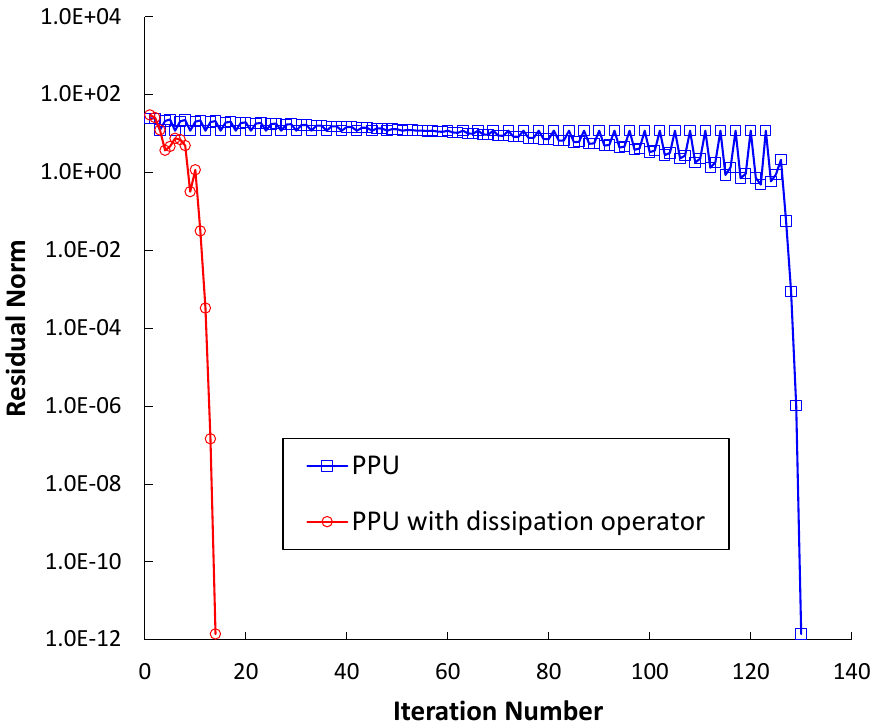}}
\\
\subfloat[Initial condition $S=0.2 $]{
\includegraphics[scale=0.7]{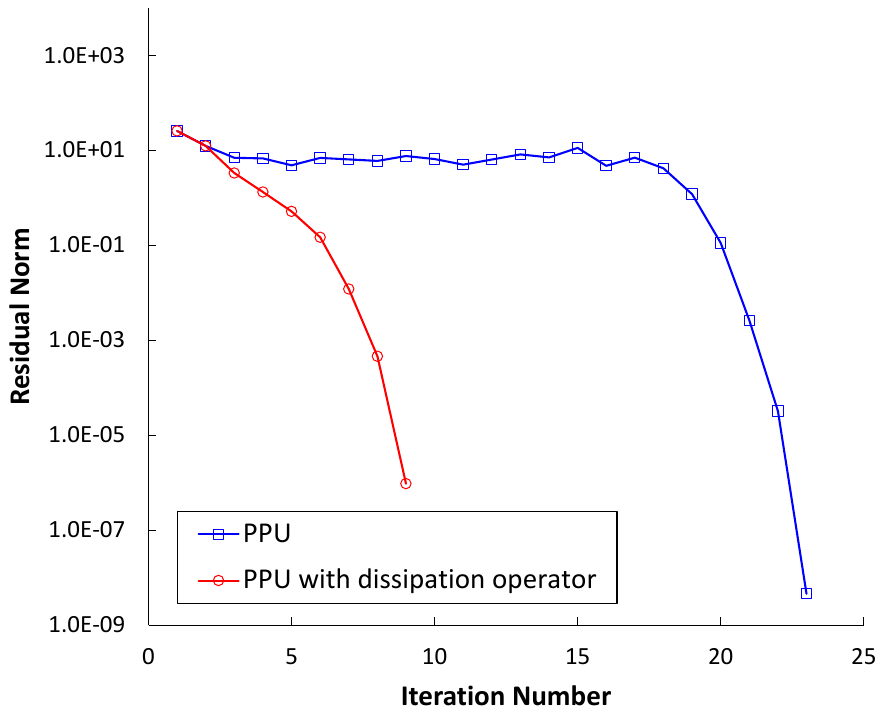}}
\caption{Residual norm versus iteration for PPU with and without the dissipation operator.}
\label{fig:resi_1}
\end{figure}

\begin{figure}[!htb]
\centering
\subfloat[Without the dissipation operator]{
\includegraphics[scale=0.7]{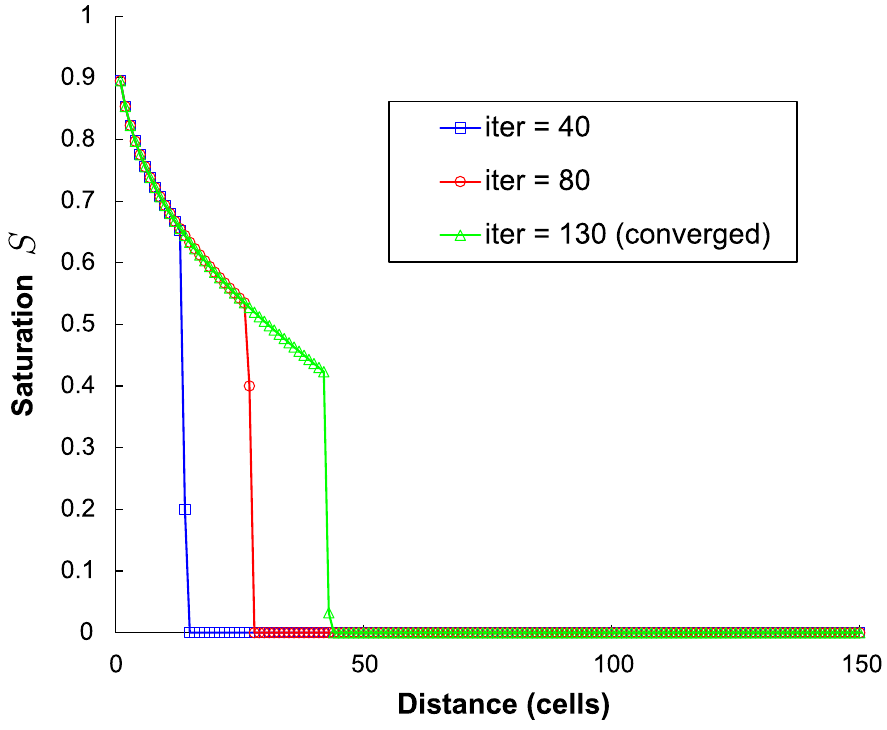}}
\\
\subfloat[With the dissipation operator]{
\includegraphics[scale=0.7]{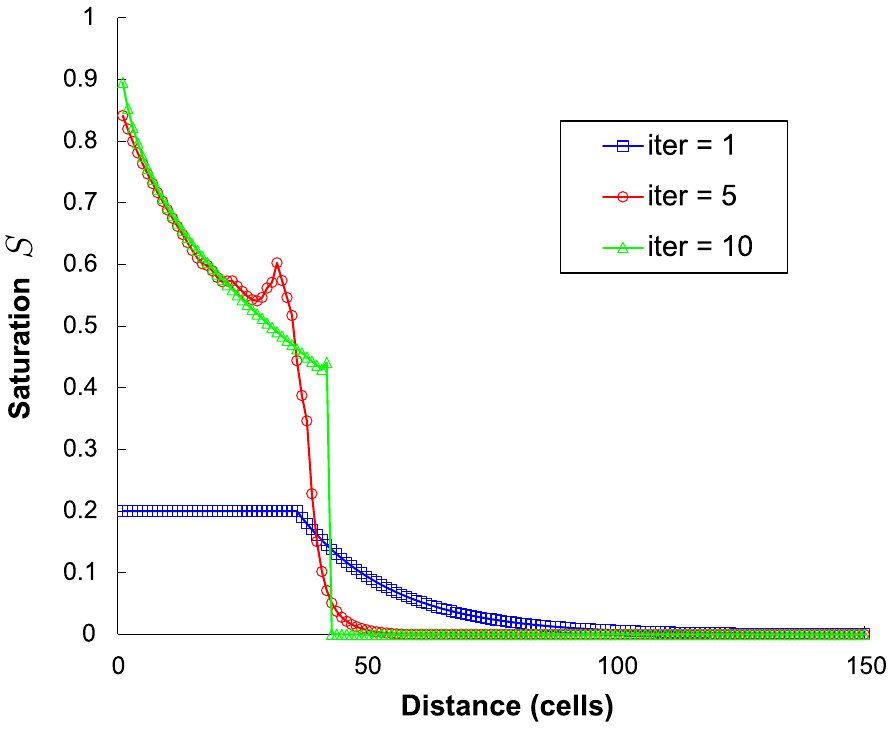}}
\caption{Saturation distributions during the iterative process for the initial condition $S=0 $.}
\label{fig:s_12}
\end{figure}

\textbf{Fig. \ref{fig:resi_1}} shows the residual norm decay versus iteration for the PPU flux - with and without the dissipation operator. $\beta $ is set to 0.14. The results for the cases with the initial conditions $S=0 $ and $S=0.2 $ are compared. From the figures, we observe a huge contrast in the iteration performance of PPU between the two initial saturations. Also a stagnation stage with oscillatory updates is clearly present in the convergence history before the residual norm starts to decrease. To investigate the reason for this type of slow convergence, we plot the saturation distribution during the iterative process for the initial condition $S=0 $ in \textbf{Fig. \ref{fig:s_12}}. Clearly, the updates have local support that 
propagates through the domain as the Newton process moves forward. 
After 130 iterations, the union of the supports of these updates finally contains the support of the state changes over the duration of the timestep, and the converged solution is obtained. This localized behavior within the Newton updates for implicit transport problems significantly degrade nonlinear convergence performance. 

\begin{figure}[!htb]
\centering
\includegraphics[scale=0.7]{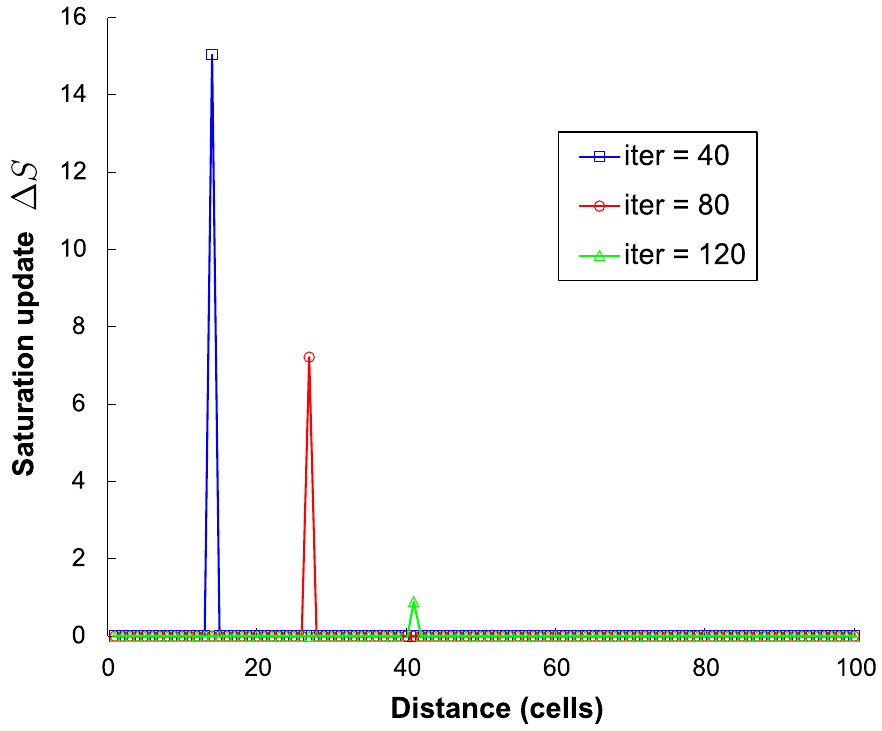}
\caption{Sequence of the Newton updates without the dissipation operator.}
\label{fig:update_1}
\end{figure}

In addition, the sequence of Newton updates without the dissipation operator is shown in \textbf{Fig. \ref{fig:update_1}}. It can be observed that there are non-monotonic `spikes' in the Newton updates. The spikes have very high saturation values, which are beyond the physical range $\left [ 0, 1 \right ]$; and hence correspond to non-physical mass accumulation. The `spikes' which are proportional to the affected domain over the timestep, propagate downstream quite slowly with the Newton iterations. 

\begin{figure}[!htb]
\centering
\includegraphics[scale=0.7]{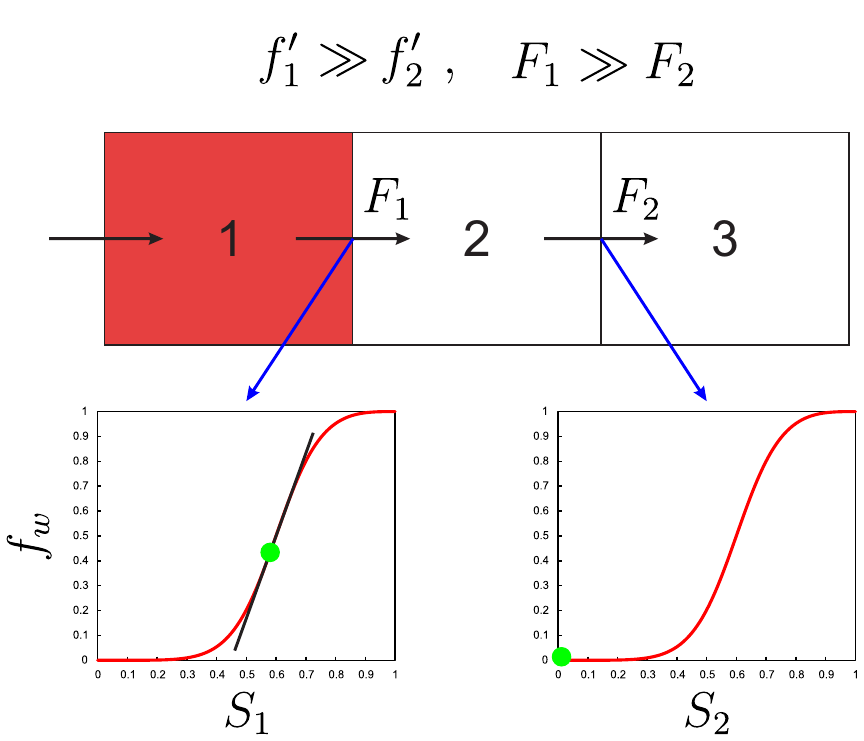}
\caption{Schematic for the three-cell example}
\label{fig:plot_3c_e}
\end{figure}

To further demonstrate the above convergence difficulty, we show a three-cell example in \textbf{Fig. \ref{fig:plot_3c_e}}. Consider a saturation front traveling from the left to the right. The propagation at the leading edge (between Cell 1 and 2), where the injected fluid is invading Cell 2 that is fully saturated with the resident fluid ($S=0$), is constrained by the zero mobility of the invading phase in that cell (Wang 2012). For each cell, the in-flux comes from its upwind neighbor. The out-flux $F_2$ from Cell 2 being invaded by the injected fluid is much less than the in-flux $F_1$ from the upstream Cell 1. We can also see that at $S=0$, the slope of the flux function, i.e., $(df/dS)$, is zero. For a given balance of forces (i.e., viscous, gravitational, and capillary), the shape of the flux function is determined primarily by the relative permeability curves. Note that the slope of the flux function is the speed of the saturation wave. In Appendix A, we analyze the wave speeds for linear relative permeability curves and more general nonlinear functions. The analysis reveals the fact that the wave speed coincides with the first-order derivative in the Newton method; thus low wave speed around the saturation front will result in small value of the Newton update. Therefore using $S = 0$ everywhere as the initial guess for the Newton solver, it is not possible to invade two successive cells in a single Newton update. This explains the localized behavior as well as the spikes associated with the non-physical mass accumulation within the Newton updates.

Compared with the standard Newton method, the homotopy continuation based on the dissipation operator leads to dramatic improvement in the convergence rate. \textbf{Fig. \ref{fig:s_12}}(b) illustrates that the saturation wave can properly spread into the domain, and the update profile is monotonic for the first Newton iteration. Therefore, there are no spikes due to the non-physical mass accumulation. During the subsequent iterations, the saturation profile gets sharpened and converges quickly to the desired solution.
In Appendix B, the convergence ratios of a single-cell problem are analyzed, showing the favorable convergence property of the dissipation operator.

\begin{figure}[!htb]
\centering
\subfloat[Initial condition $S=0 $]{
\includegraphics[scale=0.7]{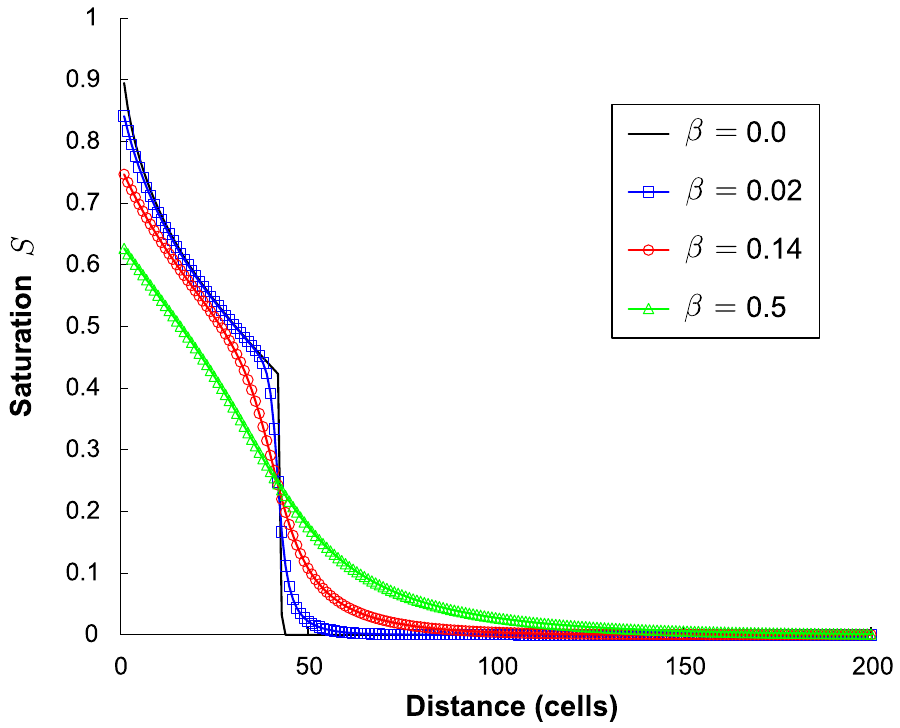}}
\\
\subfloat[Initial condition $S_{\textrm{lower}} = 0 $ and $S_{\textrm{upper}} = 1 $]{
\includegraphics[scale=0.7]{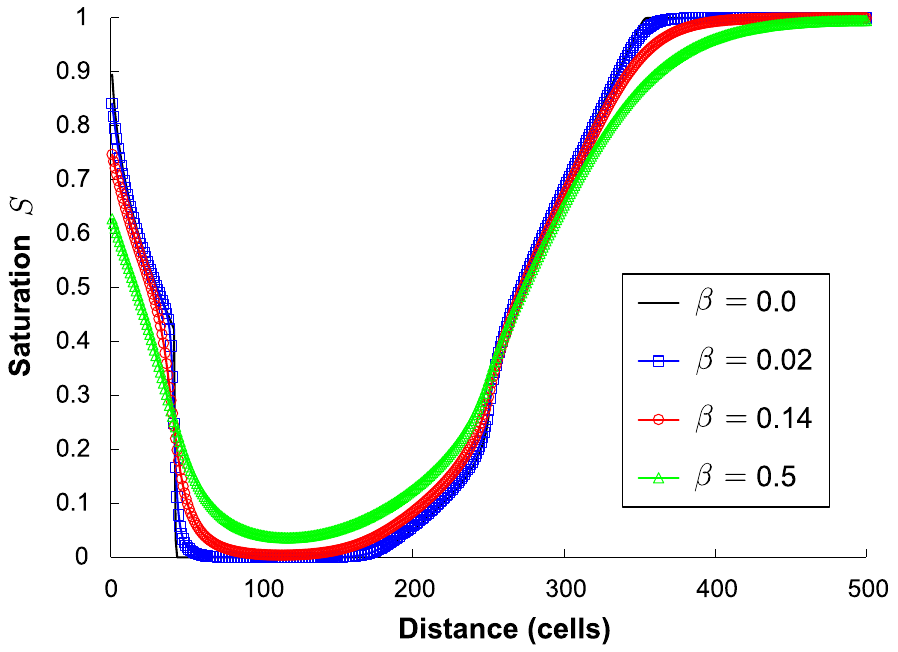}}
\caption{Saturation distributions with fixed continuation parameter $\kappa = 1 $}
\label{fig:S_profile}
\end{figure}

Now we examine the impact of the dissipation coefficient $\beta$ on the solutions of the scalar nonlinear transport problem. The initial condition with $S_{\textrm{lower}} = 0 $ and $S_{\textrm{upper}} = 1 $ (oil in the lower half of the domain, and water in the upper half) is also considered. The saturation distributions for different values of $\beta$ with fixed continuation parameter $\kappa = 1 $ (without following the homotopy path) are plotted in \textbf{Fig. \ref{fig:S_profile}}. It can be seen that increasing $\beta$ has the qualitative effect of `smearing' the numerical solutions. Essentially, with the artificial dissipation term, 
the hyperbolic equation is transformed into an elliptic one (Brown and Zingg 2017).

\subsection{Adaptive dissipation coefficient}

The previous results show that the DBC method is effective and can significantly improve the nonlinear convergence for the scalar transport problem. The method can overcome the convergence difficulties caused by the low wave speed of the saturation front. For practical applications, the mass associated with the timestep needs to be properly spread downstream to ensure that the non-physical mass accumulation does not appear; at the same time the mass should not be spread too far away from the desired solution. Here, we propose an adaptive strategy to determine the value of the dissipation coefficient $\beta$. 
To derive the adaptive strategy, we first introduce the interface flux in a central scheme, which can be written in the form of
\begin{equation} 
F_{ij} = \frac{1}{2}\left [ f(S_i) + f(S_j) \right ] + \frac{1}{2} \gamma \left ( S_i - S_j \right )  
\end{equation}
The first term (average flux) represents the central discretization of the flux, and the second term is the diffusive flux. In the Local Lax-Friedrichs (LLF) flux, $\gamma$ is taken as the maximum absolute value of the wave speed between the left and right cell states, that is,
\begin{equation} 
\gamma = \underset{s \in \left [ S_i, S_j \right ]}{\mathrm{max}}\left | {f}'(s) \right |
\end{equation}
Note that the affected domain over a timestep during a hyperbolic transport process is proportional to the physically relevant speed of wave propagation. Therefore, $\beta$ should be derived in a CFL-like formula as
\begin{equation} 
\beta = \omega \frac{u_T \Delta t}{\phi \Delta x} \mathrm{max}\left | {f}'(s) \right |
\label{Eq:ada_sm}
\end{equation}
where $\omega$ is a tunable coefficient. It is expected that a suitable value of $\omega $ can be chosen for a target class of problem. We find out that $\omega = 2.0e-3 $ is effective for the 1D scalar transport presented in this work. Compared to the LLF flux, the maximum derivative of the analytical flux, instead of the maximum within the local cell states, is adopted in the adaptive formula.

\subsection{Viscous-gravitational case}

\begin{figure}[!htb]
\centering
\includegraphics[scale=0.6]{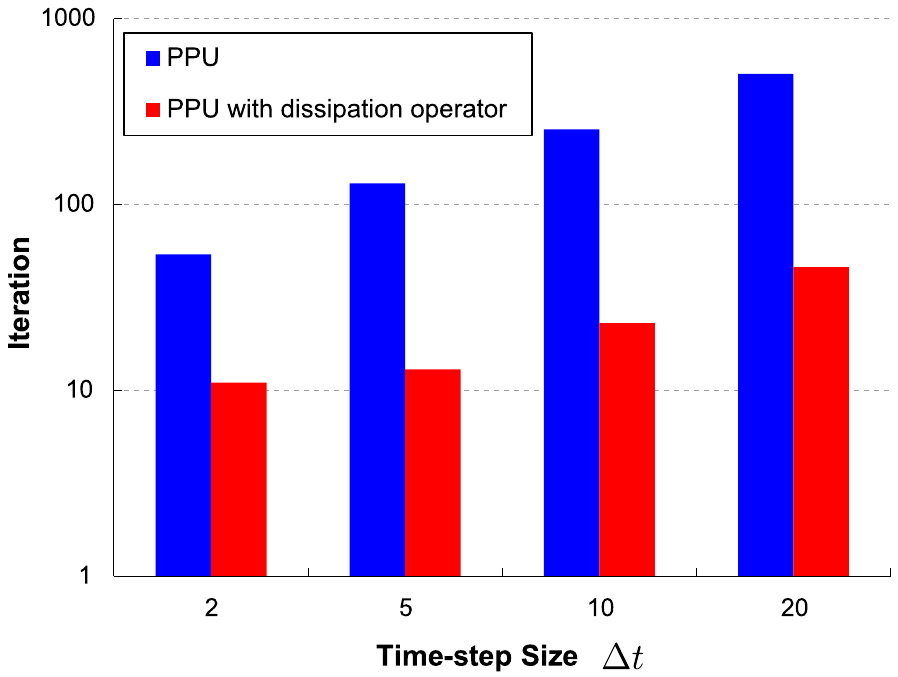}
\caption{Nonlinear iteration performance for PPU with different timestep sizes.}
\label{fig:compare_vg_1}
\end{figure}

We test the homotopy continuation method based on the adaptive dissipation strategy for cases with combined viscous and gravitational forces. 
We still consider the previous simulation example. The dissipation coefficient $\beta $ for $\Delta t = 5.0 $ is computed as 0.1389. For the initial condition $S=0 $, the nonlinear iteration performances of the PPU scheme with different timestep sizes are shown in \textbf{Fig. \ref{fig:compare_vg_1}}. We can see that the continuation method has superior convergence property. For the standard Newton method, the number of nonlinear iterations is proportional to timestep size because the solution front propagates a longer distance with a larger timestep size. 

\begin{figure}[!htb]
\centering
\includegraphics[scale=0.6]{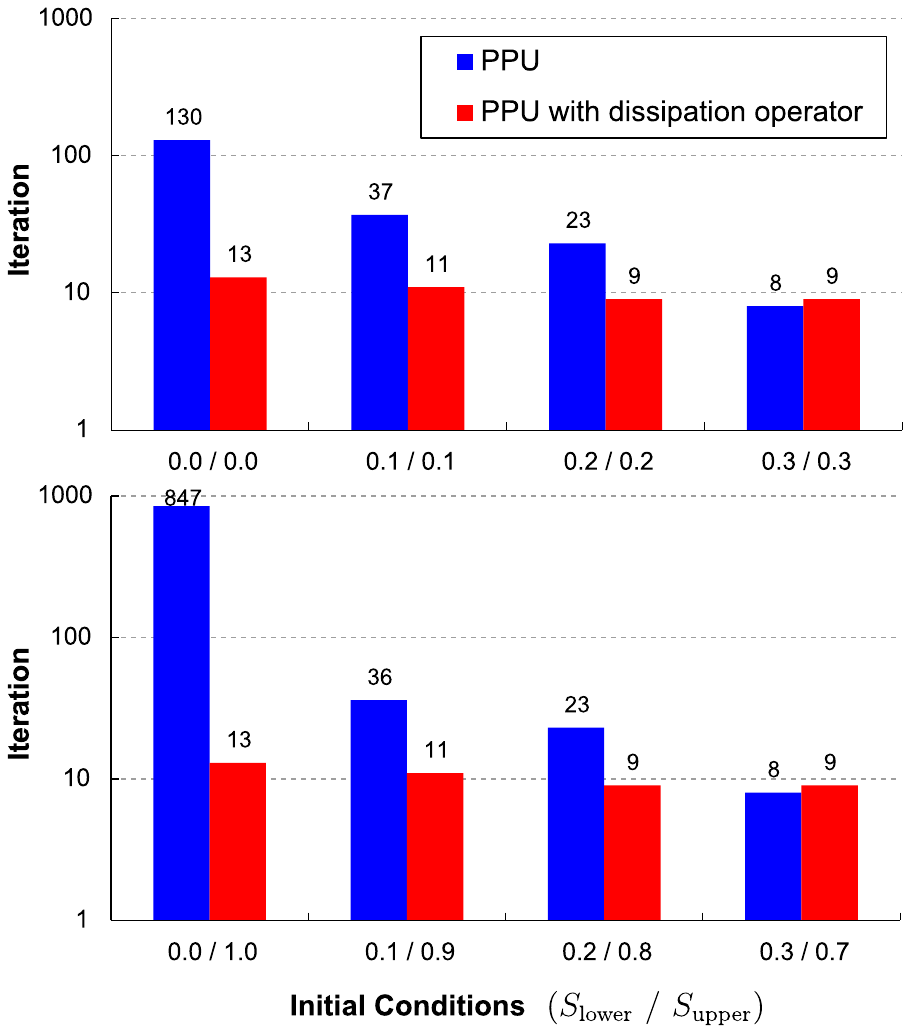}
\caption{Nonlinear iteration performance with different initial conditions.}
\label{fig:compare_vg_2}
\end{figure}

We divide the model domain into two equal parts (lower and upper parts), and we assign different initial saturations to each part. The maximum allowable number of Newton iterations per timestep is set to 1000. The nonlinear iterations for the PPU scheme with different initial conditions are shown in \textbf{Fig.\ \ref{fig:compare_vg_2}}. In contrast to the poor convergence performance of the standard Newton method, the continuation method achieves reduction in the total iterations by more than an order of magnitude for most cases; and that results in a corresponding reduction in the overall computational cost. 

We observe that the dissipation operator can always reduce the required number of iterations, to varying degrees, even for the initial conditions that are some distance away from the residual saturations. This is due to the hyperbolic nature of the transport problem, 
which is essentially an information propagation process. Therefore, the artificial dissipation can help with spreading mass in the affected domain over the timestep and lead to faster nonlinear convergence.

\subsection{Viscous-dominated case}

The flux function is the main source of nonlinear convergence difficulty for the Newton method. The details of nonmonotonicity and nonconvexity of the flux function depend on the relative-permeability relations, viscosity ratio, and the balance between the viscous and gravitational forces. It can be challenging to develop physics-based nonlinear solvers because the coupled nonlinear conservation equations governing multiphase flow and transport in heterogeneous porous media are quite difficult to analyze.

To demonstrate the effectiveness of the new nonlinear solver for the transport problems, we consider three types of the relative-permeability curves
\\
1. Corey-type quadratic:
\begin{equation} 
k_{rw} = S^2, \qquad k_{ro} = (1-S)^2 
\end{equation}
\\
2. Corey-type cubic:
\begin{equation} 
k_{rw} = S^3, \qquad k_{ro} = (1-S)^3 
\end{equation}
\\
3. Brooks-Corey:
\begin{equation} 
k_{rw} = S^4, \qquad k_{ro} = \left ( 1-S \right )^2 \left ( 1-S^2 \right )
\end{equation}

\begin{figure}[!htb]
\centering
\includegraphics[scale=0.6]{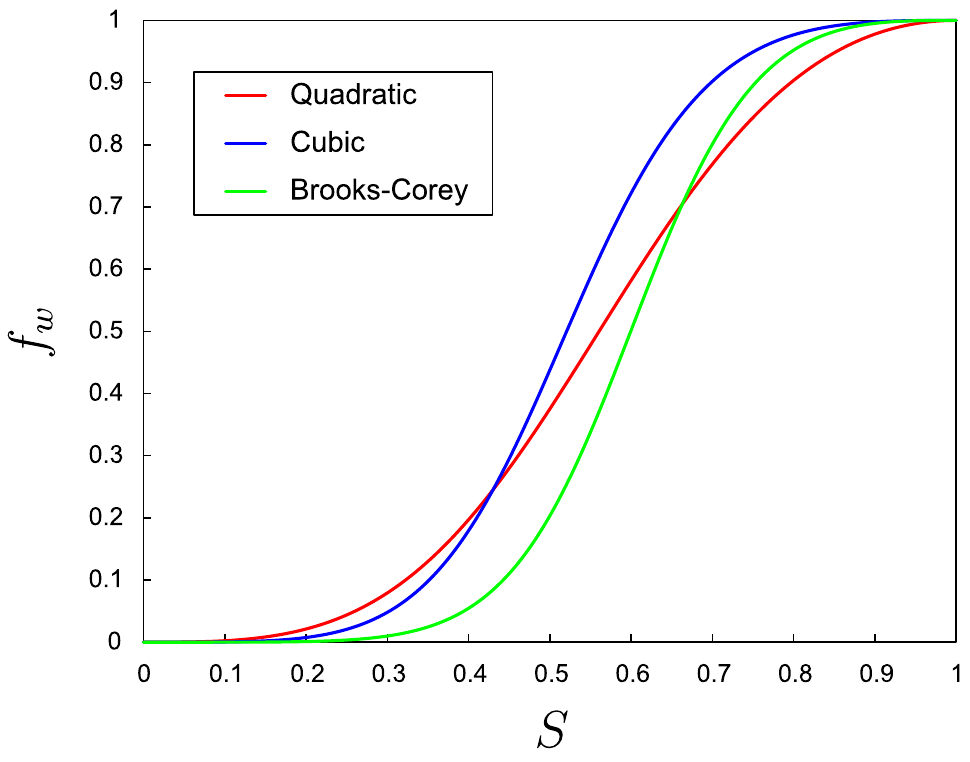}
\caption{Fractional flow curves for different relative permeability functions.}
\label{fig:vis_flux}
\end{figure}

\begin{figure}[!htb]
\centering
\includegraphics[scale=0.6]{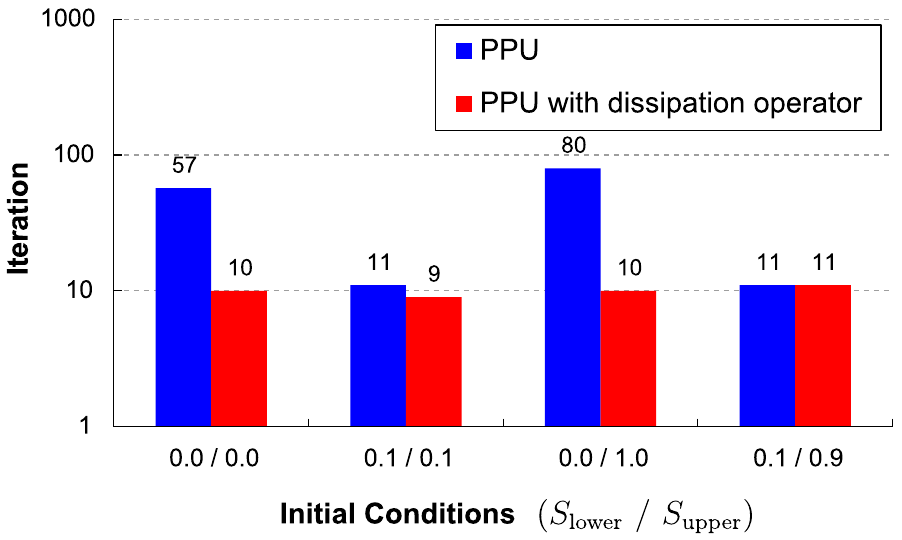}
\caption{Nonlinear iteration performance for quadratic $k_{r}$}
\label{fig:compare_V_1}
\end{figure}

\begin{figure}[!htb]
\centering
\includegraphics[scale=0.6]{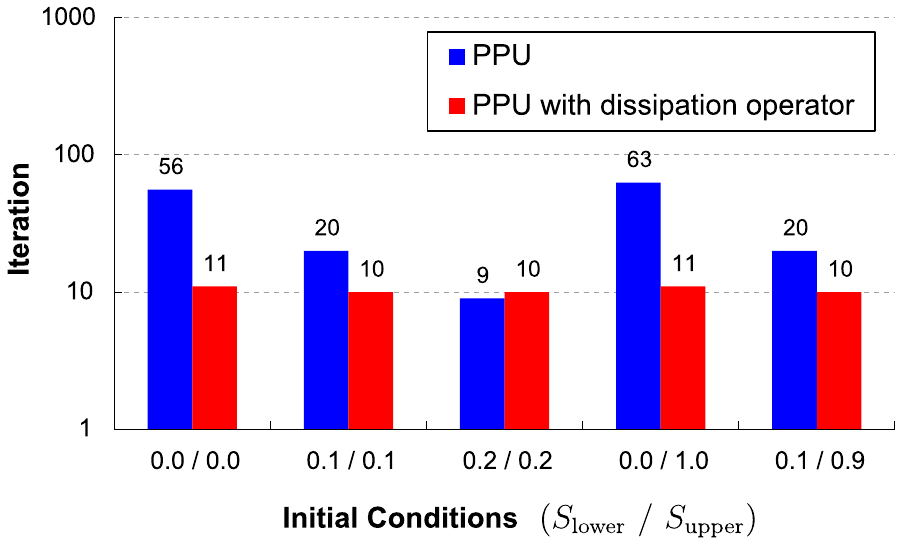}
\caption{Nonlinear iteration performance for cubic $k_{r}$}
\label{fig:compare_V_2}
\end{figure}

\begin{figure}[!htb]
\centering
\includegraphics[scale=0.6]{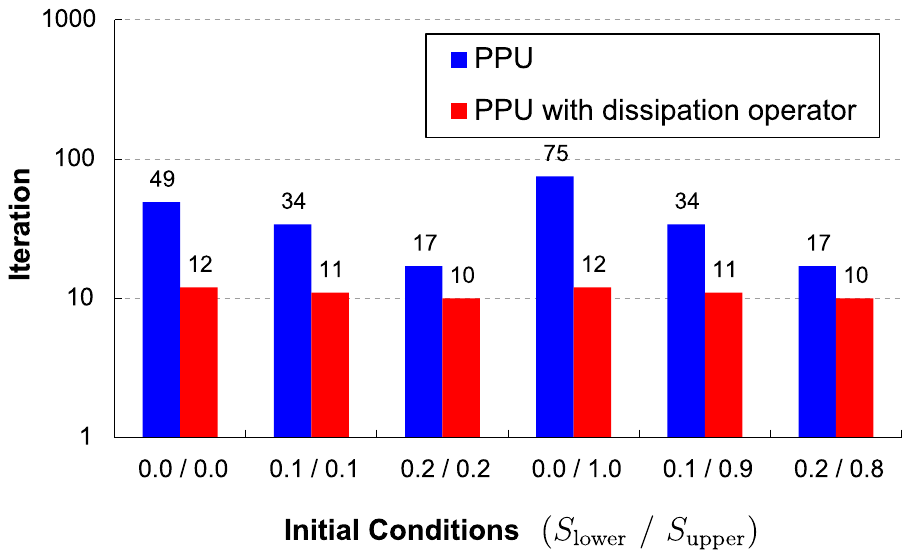}
\caption{Nonlinear iteration performance for Brooks-Corey $k_{r}$}
\label{fig:compare_V_3}
\end{figure}

We present a viscous-dominated case with $\Delta t = 5.0 $, $u_T = 0.01 $, $M = 1 $ and $C_g = 0.01 $ ($\textrm{CFL} \approx $ 52). In \textbf{Fig. \ref{fig:vis_flux}}, we plot the fractional-flow curves for the relative permeability functions listed above.
The PPU flux is employed in the numerical examples. The nonlinear iteration performance for the three types of relative-permeability curves with different initial conditions is shown in \textbf{Figs. \ref{fig:compare_V_1}}, \textbf{\ref{fig:compare_V_2}} and \textbf{\ref{fig:compare_V_3}}.

\begin{figure}[!htb]
\centering
\includegraphics[scale=0.6]{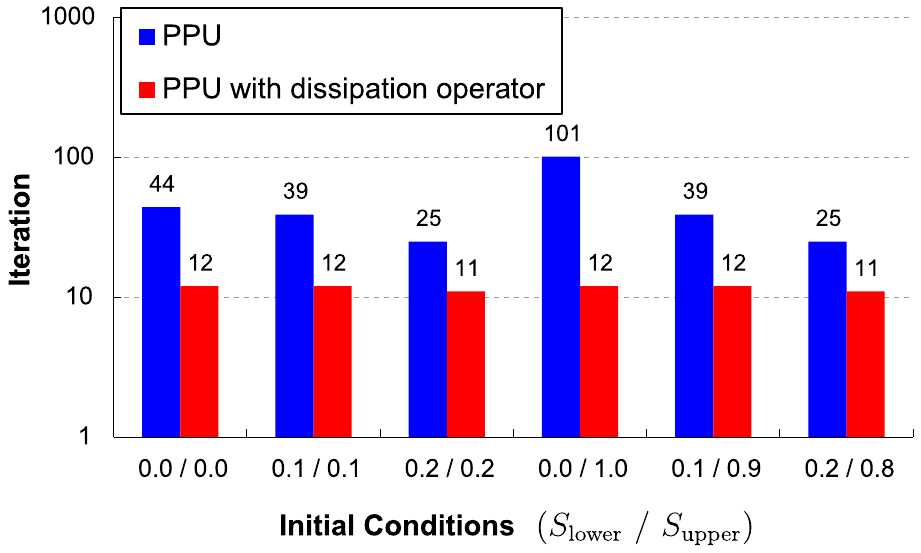}
\caption{Nonlinear iteration performance for Brooks-Corey $k_{r}$ with $M = 0.2 $}
\label{fig:compare_V_4}
\end{figure}

\begin{figure}[!htb]
\centering
\includegraphics[scale=0.7]{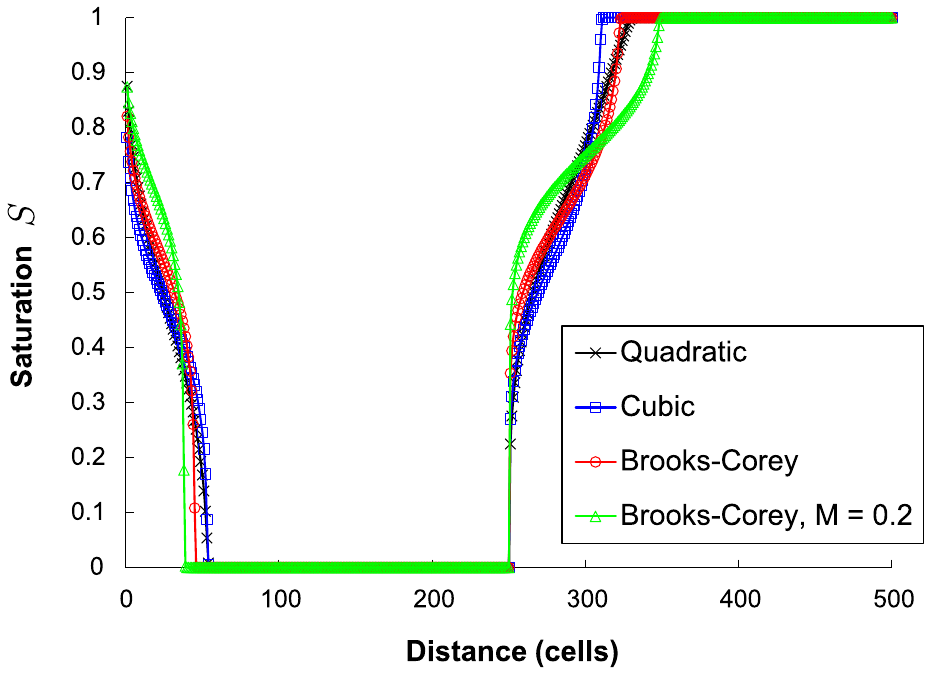}
\caption{Saturation distributions for the four cases with the initial condition $S_{\textrm{lower}} = 0 $ and $S_{\textrm{upper}} = 1 $}
\label{fig:compare_Vis_S}
\end{figure}

We also test the case with $M = 0.2 $ (the other parameters remain unchanged) for Brooks-Corey $k_{r}$ ($\textrm{CFL} \approx $ 97). The results are presented in \textbf{Fig. \ref{fig:compare_V_4}}. The saturation distributions for the above four cases with the initial condition $S_{\textrm{lower}} = 0 $ and $S_{\textrm{upper}} = 1 $ are plotted in \textbf{Fig. \ref{fig:compare_Vis_S}}. 
From the figure and iteration performance, we observe that the nonlinear convergence of the standard Newton method is closely associated with the size of the domain affected by wave propagation within a timestep. With larger size of the affected domain, the saturation wave needs to propagate over a longer distance until it reaches the front of the converged solution. Correspondingly, we expect more nonlinear iterations required by the scenario with larger affected domain. For the initial conditions that are some distance away from the residual saturations, we can conclude that sharper solution front (shock) leads to more convergence difficulty. Note that even in such cases, the dissipation operator can reduce the number of iterations significantly; 
thus, yielding great potential in the saving of computational time.

\subsection{Buoyancy-dominated case}

\begin{figure}[!htb]
\centering
\includegraphics[scale=0.6]{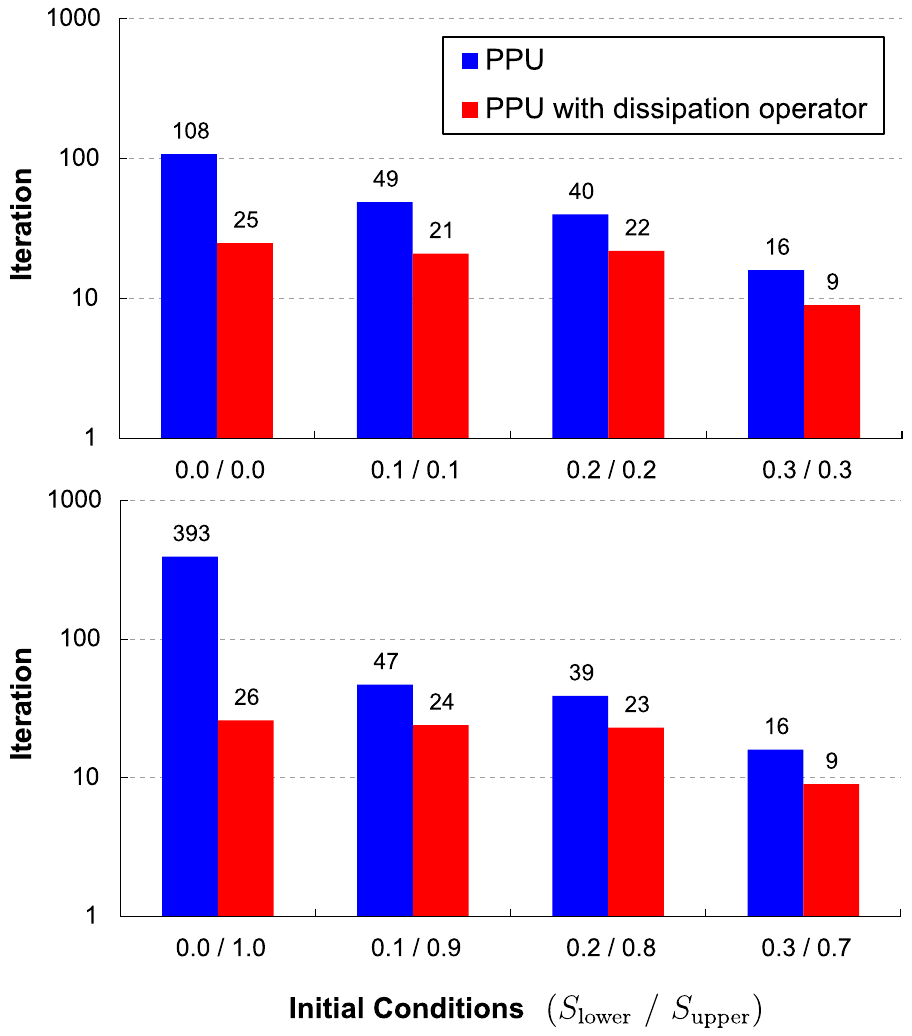}
\caption{Nonlinear iteration performance for PPU with different initial conditions in the buoyancy-dominated case.}
\label{fig:compare_G_1}
\end{figure}

We present a buoyancy-dominated case with $\Delta t = 5.0 $, $u_T = 0.01 $, $M = 5 $ and $C_g = 0.06 $ ($\textrm{CFL} \approx $ 118). The PPU scheme with quadratic $k_{r}$ is employed for the simulation examples. The nonlinear iteration performance for different initial conditions are summarized in \textbf{Fig. \ref{fig:compare_G_1}}.\ As the figure shows, significant convergence difficulties are encountered in solving the transport problem with strong buoyancy, and the DBC method can greatly improve the overall convergence behavior. Moreover, a reasonable reduction in the number of iterations can be achieved for the initial conditions that are far away from the residual saturations. 
This is mainly due to the nonlinear complexity in the numerical flux with gravitational force.

\begin{figure}[!htb]
\centering
\includegraphics[scale=0.6]{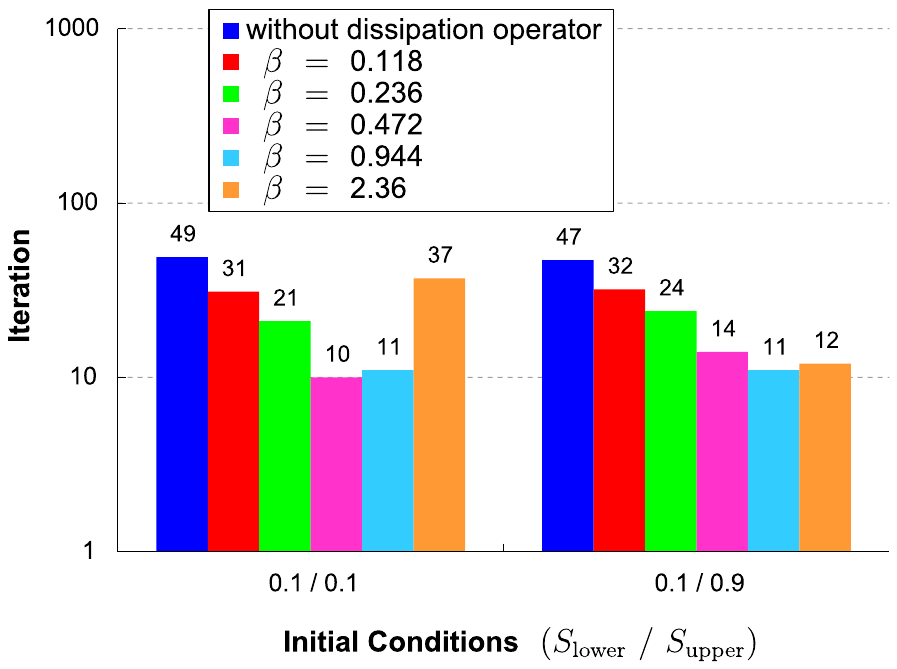}
\caption{Nonlinear iteration performance for PPU with different $\beta$ and initial conditions.}
\label{fig:compare_G_2}
\end{figure}

The impact of the dissipation coefficient $\beta$ on the nonlinear convergence is examined in \textbf{Fig. \ref{fig:compare_G_2}}. The results demonstrate that the solution performance is not very sensitive to the value of $\beta$. For practical applications, a variation of $\beta$ within an order of magnitude will not cause large negative impact on the effectiveness of the dissipation operator. Superior convergence performance can be achieved as long as $\beta$ is in the value range 
approximated by the adaptive formula. $\beta$ can be determined $\textit{a-priori}$ before each timestep during a simulation. 

\subsection{Capillary case}

We examine the effects of the capillary force on the nonlinear convergence of the transport problems. The model parameters are the same as the one used for the viscous-gravitational case. The PPU flux is utilized for all the tests. The following capillary-pressure curve is 
\begin{equation} 
P_c(S) = P_{c,e} S^{-0.5}
\end{equation}
where $P_{c,e}$ is the capillary entry pressure, which is taken as 0.1 bar. The characteristic length ($L$) in Eq.\ (\ref{Eq:f_c_11}) equals the length of a cell, and the characteristic capillary pressure ($\bar{P_c}$) is 0.1 bar.

\begin{figure}[!htb]
\centering
\includegraphics[scale=0.6]{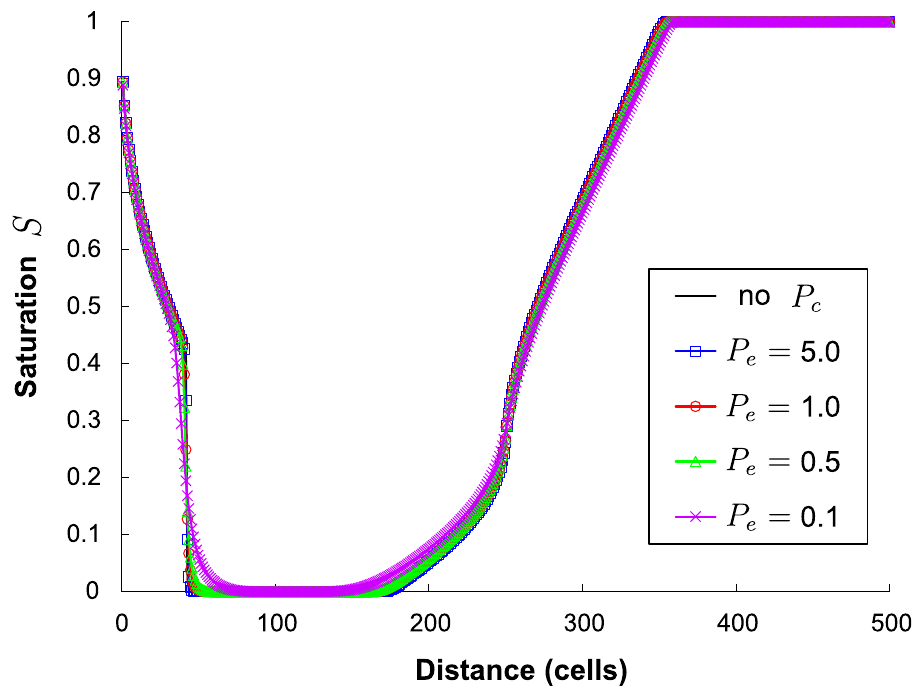}
\caption{Saturation distributions for (PPU + capillarity) with different Peclet number $P_e$}
\label{fig:compare_Pc}
\end{figure}

The initial condition with $S_{\textrm{lower}} = 0 $ and $S_{\textrm{upper}} = 1 $ (oil on the lower half of the domain and water above) is specified.\ The saturation distributions of the solutions for different Peclet number, $P_e$, are plotted in \textbf{Fig. \ref{fig:compare_Pc}}.\ As we can see, the physical diffusion induced by capillarity has a similar effect with the artificial dissipation on the solution profile. Smaller value of $P_e$ indicates stronger capillary force, resulting in larger dissipation effect around the saturation fronts.

\begin{figure}[!htb]
\centering
\subfloat[Initial condition $\left ( \ 0.0 \ / \ 1.0 \ \right )$]{
\includegraphics[scale=0.7]{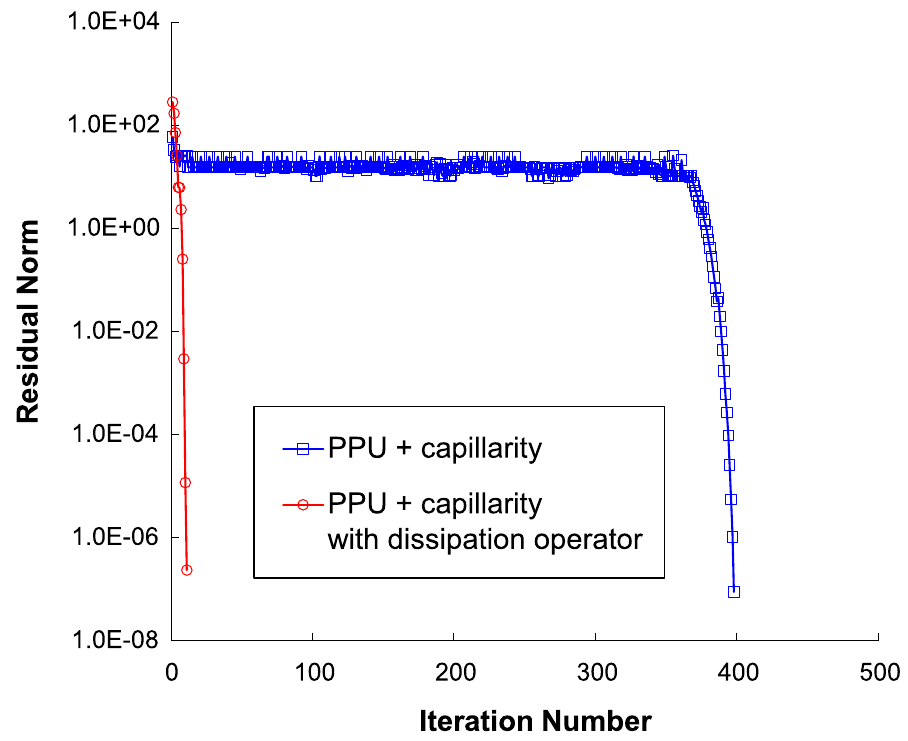}}
\\
\subfloat[Initial condition $\left ( \ 0.2 \ / \ 0.2 \ \right )$]{
\includegraphics[scale=0.7]{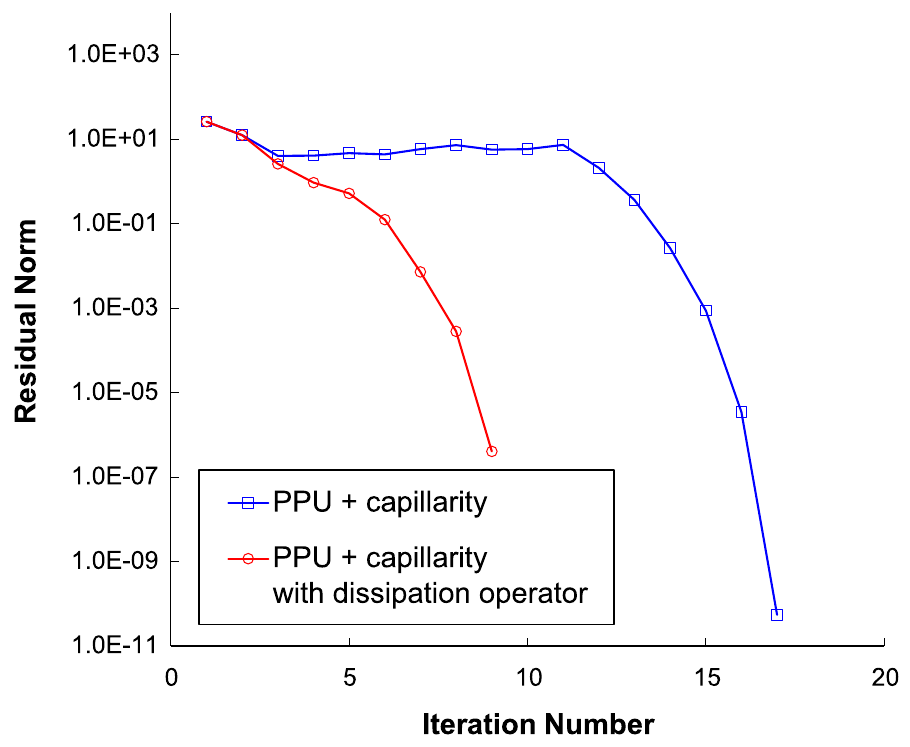}}
\caption{Residual norm versus iteration for (PPU + capillarity) with $P_e = 1.0 $}
\label{fig:resi_Pc}
\end{figure}

\begin{figure}[!htb]
\centering
\includegraphics[scale=0.6]{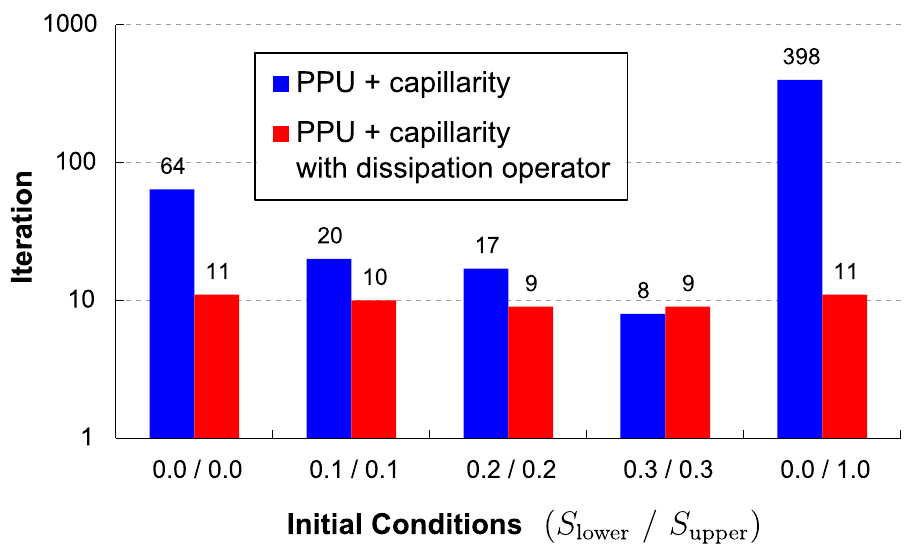}
\caption{Nonlinear iteration performance for (PPU + capillarity) with $P_e = 1.0 $ and different initial conditions.}
\label{fig:compare_Pc_1}
\end{figure}

\begin{figure}[!htb]
\centering
\includegraphics[scale=0.6]{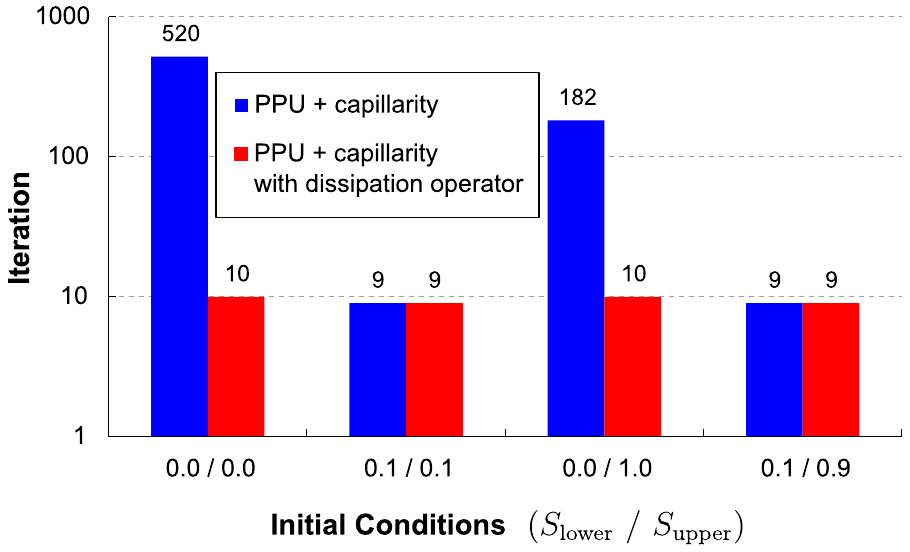}
\caption{Nonlinear iteration performance for (PPU + capillarity) with $P_e = 0.1 $ and different initial conditions.}
\label{fig:compare_Pc_2}
\end{figure}

\textbf{Fig. \ref{fig:resi_Pc}} shows the residual norm decay versus iteration for $P_e = 1.0 $ - with and without the dissipation operator. We observe that for the moderate value of $P_e$, the reduction in residual still stagnates 
as the solver jumps between successive iterates. The nonlinear iterations for different initial conditions are summarized in \textbf{Fig. \ref{fig:compare_Pc_1}}. The nonlinear iterations for $P_e = 0.1 $ with different initial conditions are shown in \textbf{Fig. \ref{fig:compare_Pc_2}}. It can be seen that capillarity 
will not alleviate the convergence difficulty due to the low wave speed and unphysical mass accumulation for the initial condition near the residual saturation. This is because of the degenerate nature of the capillary diffusion term. In contrast to the poor performance of the standard Newton method, the DBC strategy reduces the number of Newton iterations significantly.

\begin{figure}[!htb]
\centering
\includegraphics[scale=0.6]{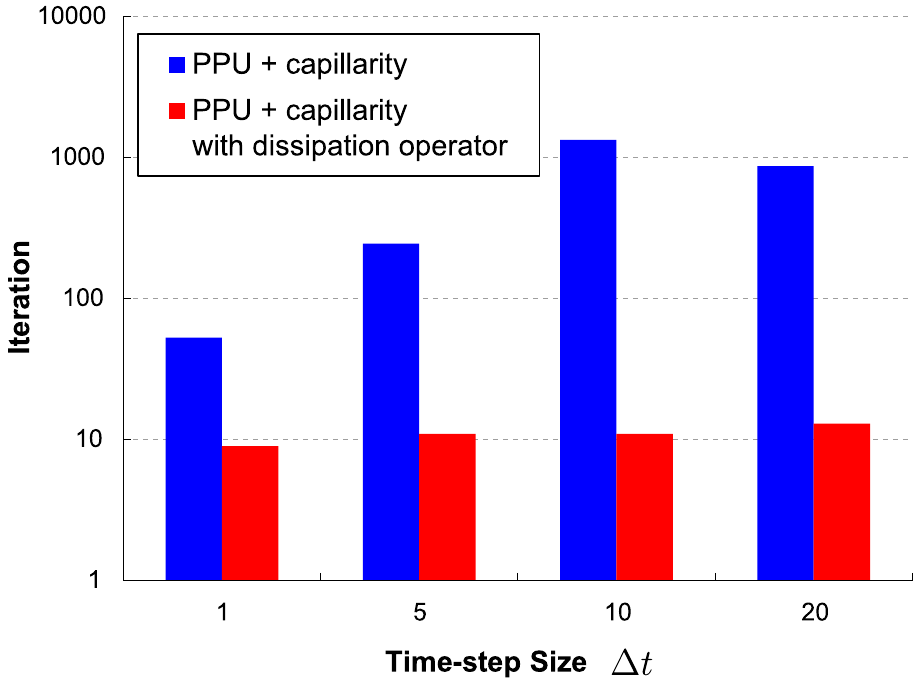}
\caption{Nonlinear iteration performance for (PPU + capillarity) with different timestep sizes.}
\label{fig:compare_gc}
\end{figure}

We consider a case of gravity segregation with capillarity ($u_T = 0 $ and $P_e = 0.5 $). For $\Delta t = 5.0 $, $\textrm{CFL} \approx $ 83. The initial condition with $S_{\textrm{lower}} = 0 $ and $S_{\textrm{upper}} = 1 $ is specified. The boundaries are closed with no source and sink. Starting from the initial condition, water will sink down, and oil will flow up. The nonlinear iteration performance of the gravitational-capillary example with different timestep sizes is shown in \textbf{Fig. \ref{fig:compare_gc}}. For this challenging case, the DBC strategy leads to a remarkable gain in speed-up.

\section{Fully-coupled flow and transport}

In this section, we intend to directly deal with the mass conservation equations for coupled flow and transport in multiple dimensions. Immiscible two-phase fluid system is considered. The mass-conservation equations in terms of the inverse formation-volume-factor (FVF) $b_\alpha$ are employed
\begin{equation} 
\frac{\partial }{\partial t }\left ( \phi b_{\alpha} S_{\alpha} \right ) + \nabla \cdot \left (b_{\alpha} \textbf{u}_{\alpha} \right ) = b_{\alpha} q_{\alpha}
\end{equation}
The phase density is evaluated through $\rho_\alpha = \rho_{\alpha, ref} b_{\alpha}$ and $\rho_{\alpha, ref}$ is reference density. We focus on the model of immiscible two-phase flow with the oil (nonwetting) and the water (wetting) phases ($\alpha=o,w$). 

An adaptive strategy for the optimum dissipation coefficient $\beta$ is derived based on the mass conservation equations, which are different from the fractional-flow formulation with separate pressure (elliptic) and saturation (hyperbolic) equations. 

\subsection{Adaptive dissipation coefficient}

The sequential-implicit method (SIM) is a popular solution strategy to handle coupled flow and transport in porous media. For each timestep in SIM, there are two loops performed in sequence: one for the pressure (total velocity) and one for the saturation. The pressure-saturation loops are wrapped with an outer loop. For each outer loop iteration, the computations proceed as follows: solve for the pressure field iteratively to a certain tolerance and update the total velocity; then compute the saturation iteratively. The adaptive formula (\ref{Eq:ada_sm}) proposed previously for $\beta$ can be straightforwardly computed for each cell interface within the SIM solution framework, because $u_T$ is assumed to be fixed during the saturation updates.

The total-velocity is generally a function of space and time in multiple dimensions. The dissipation flux from Eq. (\ref{Eq:d_scalar}) can be readily employed in the fully-coupled problem. An adaptive formula for $\beta$ will be derived under the fractional-flow formulation. The total velocity discretization is given by
\begin{equation} 
u_{T,ij} = \sum_{m} T_{ij} \lambda_{m,ij} \Delta \Phi_{m,ij} = T_{ij} \lambda_{T,ij} \Delta p_{ij} + T_{ij} \sum_{m} \lambda_{m,ij} g_{m,ij} 
\end{equation}
The discrete phase flux is written as a function of the total flux
\begin{equation} 
F_{\alpha,ij} = \frac{\lambda_{\alpha,ij}}{\lambda_{T,ij}} u_{T,ij} + T_{ij} \sum_{m} \frac{\lambda_{\alpha,ij} \lambda_{m,ij}}{\lambda_{T,ij}} \left ( g_{\alpha,ij} - g_{m,ij} \right ) 
\end{equation}
where the discrete weights are $g_{\alpha,ij} = \rho_{\alpha} g \Delta h_{ij} $. 

Now the adaptive dissipation coefficient for the water flux is expressed as 
\begin{equation} 
\beta_{ij} = \omega \frac{\Delta t}{\phi_{ij} \left | \Omega_{ij} \right | } \mathrm{max}\left | {F}'_{w,ij} \right | 
\label{Eq:ada_fft}
\end{equation}
where $\beta$ is locally computed for each cell interface and $\beta_{ij}$ denotes the value for interface $(ij)$. To avoid degrading the performance of the homotopy continuation method, the total flux and the dynamic properties in the gravitational term from timestep $n$ are used for the current timestep $(n+1)$. In this way $\beta_{ij}$ is fixed during the iterative process of a timestep. We use $\omega = 1.0e-5 $ for all the following simulation cases.

\subsection{Results}

\begin{figure}[!htb]
\centering
\subfloat[Permeability (md)]{
\includegraphics[scale=0.61]{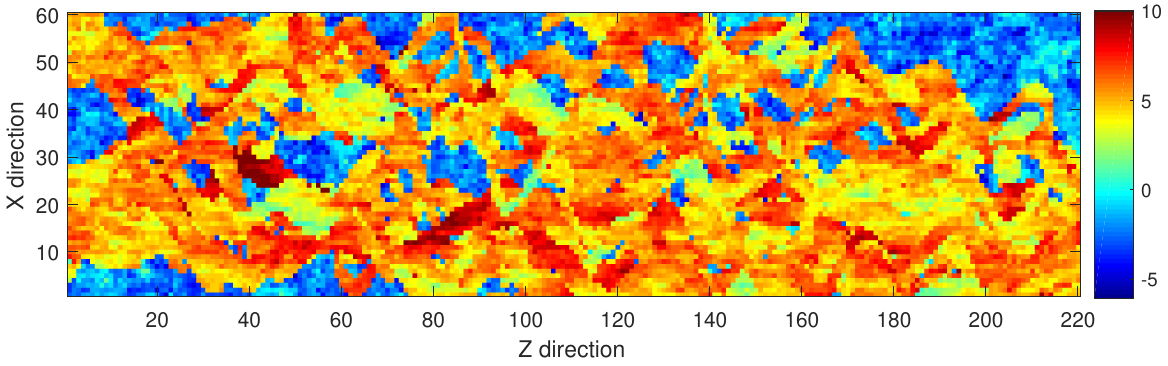}}
\\
\subfloat[Porosity]{
\includegraphics[scale=0.6]{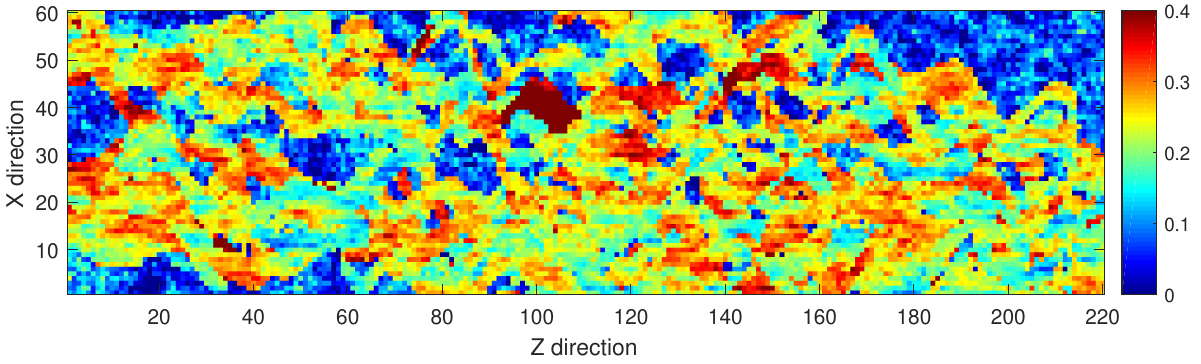}}
\caption{Permeability and porosity fields of the bottom layer in the SPE 10 model}
\label{fig:perm_poro}
\end{figure}

We demonstrate the effectiveness of the new nonlinear solver for complex heterogeneous reservoir models. The specification of the base model is shown in Table \ref{tab:specification}. The PVT properties for dead oil (PVDO) are shown in Table \ref{tab:PVDO}. Quadratic relative-permeability functions are used for the base model. The rock properties shown in \textbf{Fig. \ref{fig:perm_poro}} represent the bottom layer in the SPE~10 model. We consider the scenario with an injector at the bottom left corner and a producer at the top right. The simulation control parameters are summarized in Table \ref{tab:control}. The solutions from previous timestep $n$ are taken to be the initial guesses. A simple time stepping strategy is employed in the simulator: if the Newton method fails to converge, the timestep is reduced by half until convergence is reached; if a reduced timestep is being used and the iteration number becomes less than the optimal number, the next timestep will be doubled. In this section, a different update schedule of $\kappa$ is employed in the DBC algorithm to better vary the dissipation level. After each Newton iteration, $\kappa$ is multiplied by a constant $m = 0.2 $. The globalization stage is performed for seven iterations; after that the additional dissipation becomes negligible, and the original problem can be solved with $\kappa = 1.0e-20 $. In the DBC algorithm, the number of iterations taken during the globalization stage will not be accounted for when checking the criteria of the maximum and optimal number of iterations.

\begin{table}[!htb]
\centering
\caption{Specification of the base model}
\label{tab:specification}
\begin{tabular}{|c|c|c|}
\hline
Parameter                  &  Value          & Unit   \\ \hline
NX / NZ                    &  60 / 220       &        \\ \hline
LX / LY / LZ               &  120 / 2 / 440  & m      \\ \hline
Initial water saturation   &  0.0            &        \\ \hline
Initial pressure           &  2500           & psi    \\ \hline
Oil reference density      &  10             & lb/ft3 \\ \hline
Water reference density    &  63             & lb/ft3 \\ \hline
Rock compressibility       &  1.0E-7         & 1/psi  \\ \hline
Rock reference pressure    &  2500           & psi    \\ \hline
Water reference pressure   &  3600           & psi    \\ \hline
Water reference viscosity  &  0.2            & cP     \\ \hline
Water compressibility      &  4E-6           & 1/psi  \\ \hline
Water viscosibility Cvw    &  1.2E-6         & 1/psi  \\ \hline
Production BHP             &  1000           & psi    \\ \hline
Injection rate             &  10             & m3/D   \\ \hline
\end{tabular}
\end{table}

\begin{table}[!htb]
\centering
\caption{PVDO}
\label{tab:PVDO}
\begin{tabular}{|c|c|c|}
\hline
Pressure (psi)   &  Oil FVF  &   Oil viscosity (cP)   \\ \hline
400	  &   1.012	  &  1.16   \\ \hline
1200  &   1.004	  &  1.164  \\ \hline
2000  &   0.996	  &  1.167  \\ \hline
2800  &   0.988	  &  1.172  \\ \hline
3600  &   0.9802  &	 1.177  \\ \hline
4400  &   0.9724  &  1.181  \\ \hline
5200  &   0.9646  &	 1.185  \\ \hline
5600  &   0.9607  &  1.19   \\ \hline
\end{tabular}
\end{table}

\begin{table}[!htb]
\centering
\caption{Simulation control parameters}
\label{tab:control}
\begin{tabular}{|c|c|c|}
\hline
Parameter                 & Value         & Unit   \\ \hline
Initial timestep size    & 20          & day    \\ \hline
Total simulation time     & 300           & day    \\ \hline
Maximum timestep size    & 30           & day    \\ \hline
Maximum number of nonlinear iterations     & 20         &     \\ \hline
Optimal number of nonlinear iterations     & 7          &     \\ \hline
\end{tabular}
\end{table}

\begin{figure}[!htb]
\centering
\includegraphics[scale=0.6]{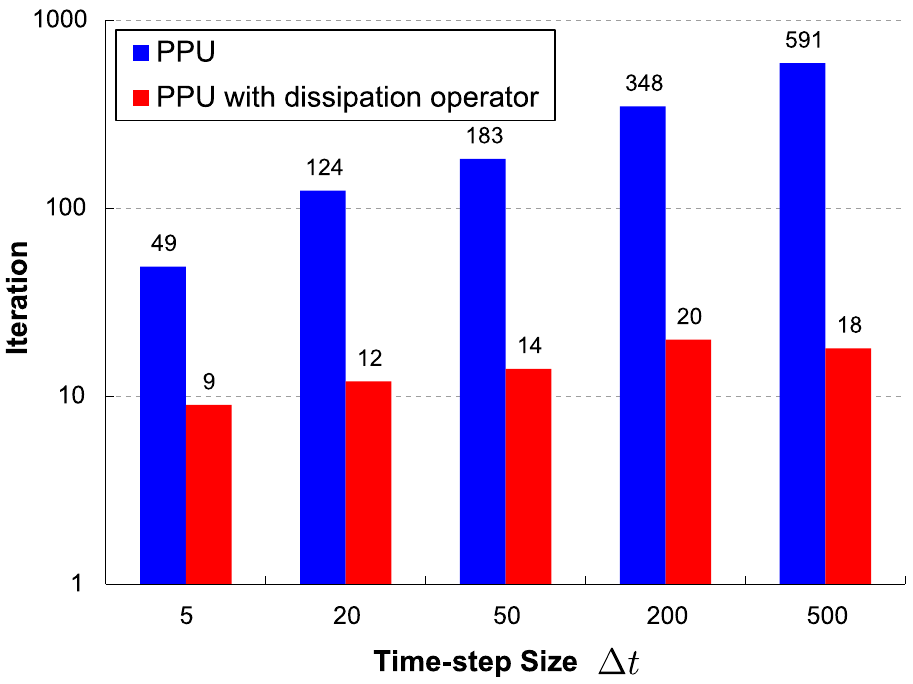}
\caption{Nonlinear iteration performance for different timestep sizes (timestep size equals total simulation time) with only viscous force.}
\label{fig:compare_2D_t}
\end{figure}

\begin{figure}[!htb]
\centering
\subfloat[50 days]{
\includegraphics[scale=0.6]{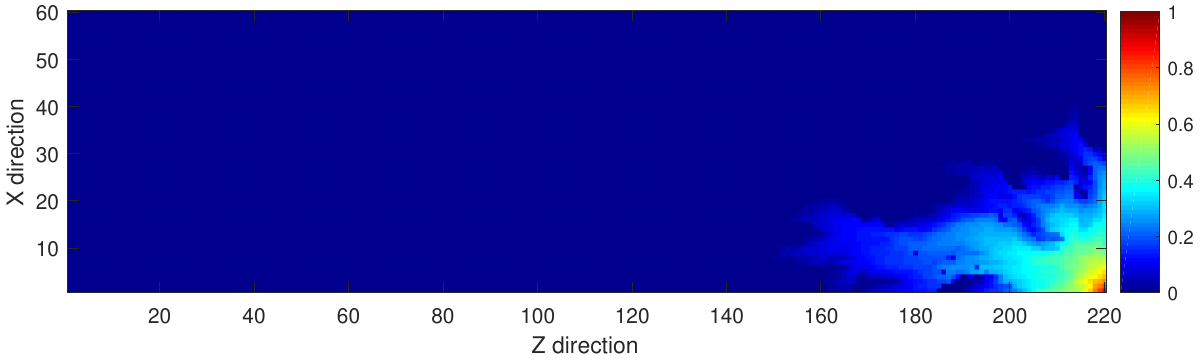}}
\\
\subfloat[200 days]{
\includegraphics[scale=0.61]{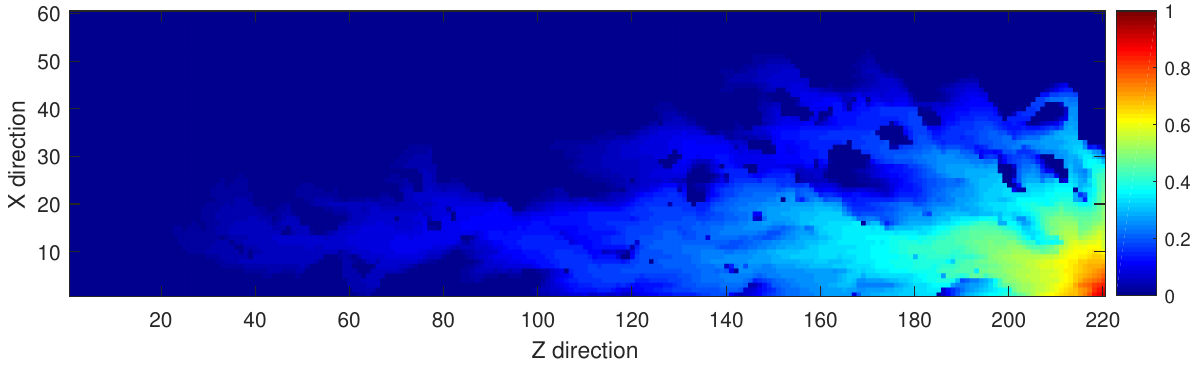}}
\\
\subfloat[500 days]{
\includegraphics[scale=0.61]{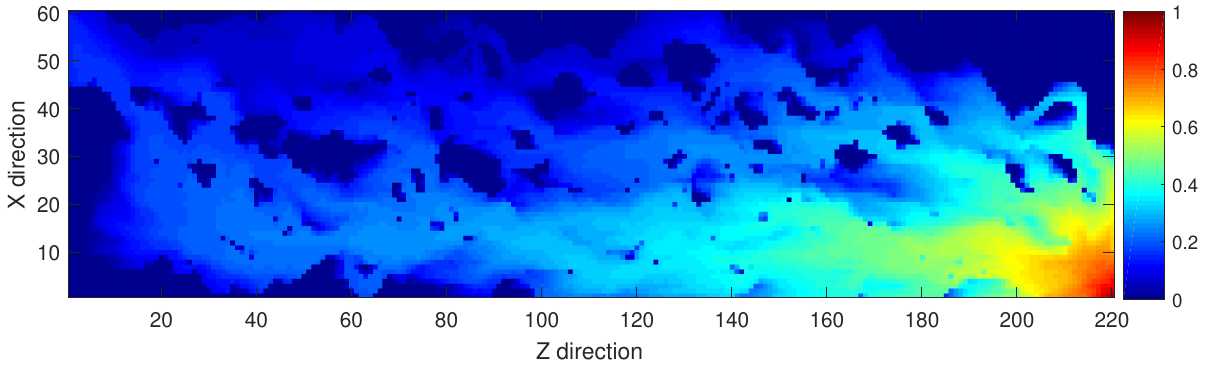}}
\caption{Water saturation profiles for different timestep sizes (timestep size equals total simulation time).}
\label{fig:2D_s_dt}
\end{figure}

We first consider the scenario with only viscous forces. The nonlinear iteration performance for different timestep sizes is summarized in \textbf{Fig. \ref{fig:compare_2D_t}}. Here, the timestep size $\Delta t$ equals the total simulation time. For $\Delta t = 5.0 $ days, the maximum $\textrm{CFL}$ number of the domain is $\textrm{CFL} \approx $ 60. The water saturation profiles are plotted in \textbf{Fig. \ref{fig:2D_s_dt}}. It can be seen that larger timestep size leads to more Newton iterations, for the reason that the solution front will propagate over a longer distance. We also observe that the iteration number of the continuation method with the dissipation operator is approximately a constant 
for the different timestep sizes.
This indicates that the developed solver is free of the CFL condition constraint. 

\begin{figure}[!htb]
\centering
\includegraphics[scale=0.6]{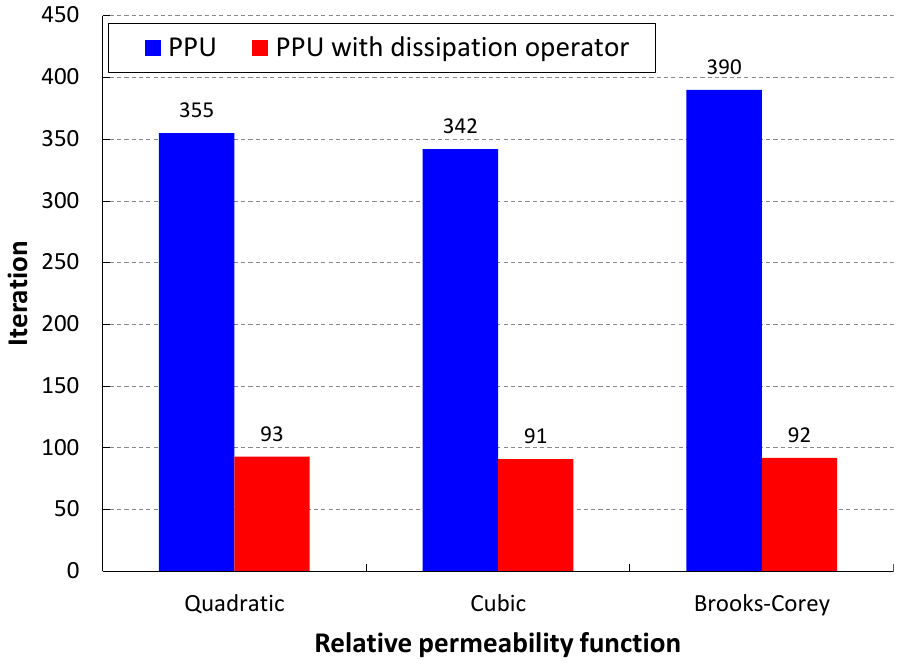}
\caption{Nonlinear iteration performance for different relative-permeability functions.}
\label{fig:compare_2D_kr}
\end{figure}

The results for the cases with different relative-permeability functions appear in \textbf{Fig. \ref{fig:compare_2D_kr}}. As can be seen, the DBC method has superior convergence performance compared to the standard Newton method, which exhibits large numbers of timestep cuts and wasted iterations. The reduction in the total Newton iterations results in a corresponding reduction in the overall computational cost.

\begin{figure}[!htb]
\centering
\includegraphics[scale=0.6]{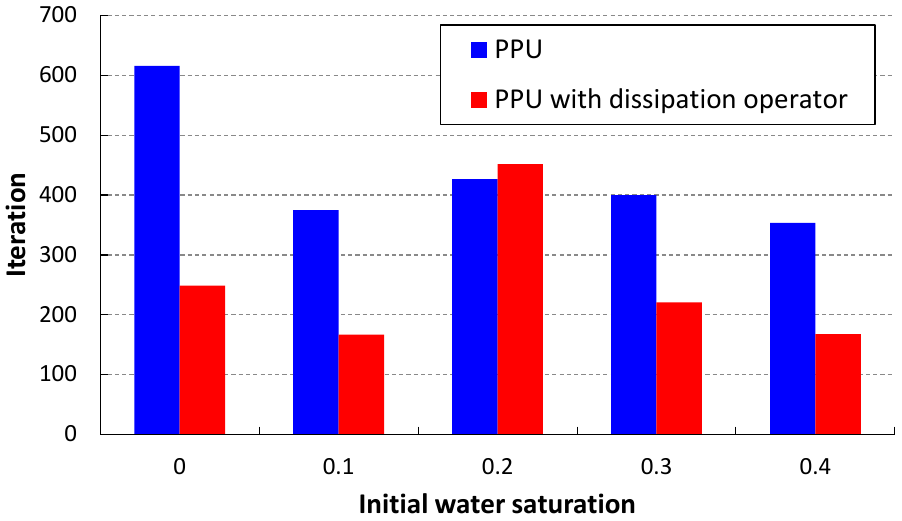}
\caption{Nonlinear iteration performance for different initial water saturations with quadratic $k_{r}$}
\label{fig:compare_2D_Q}
\end{figure}

\begin{figure}[!htb]
\centering
\includegraphics[scale=0.6]{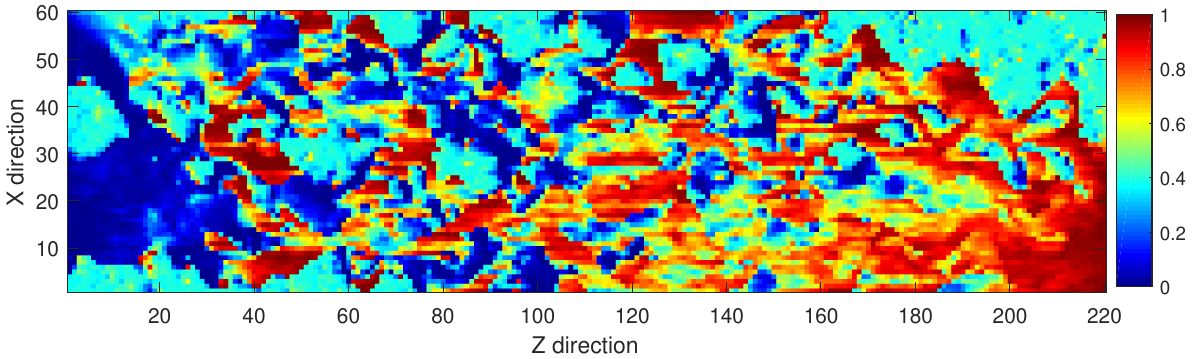}
\caption{Water saturation profile for the initial water saturation 0.4 }
\label{fig:compare_2D_Sw}
\end{figure}

\begin{figure}[!htb]
\centering
\includegraphics[scale=0.6]{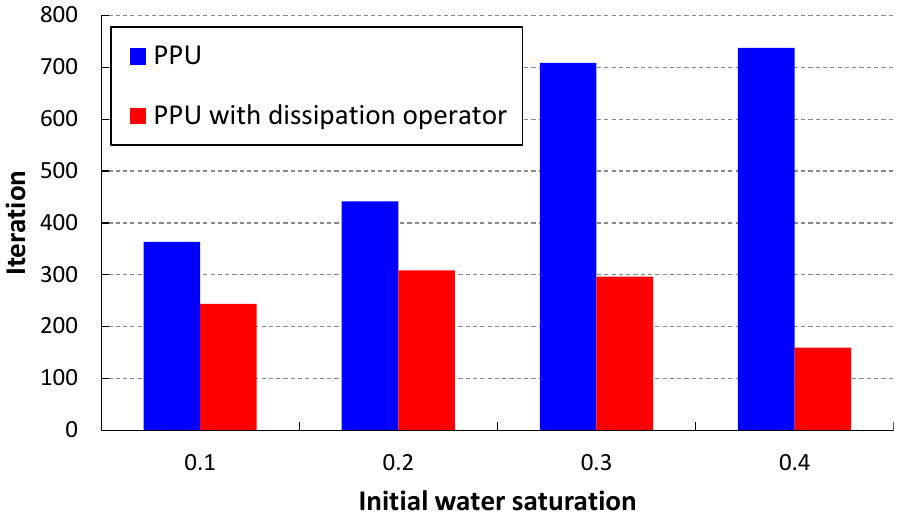}
\caption{Nonlinear iteration performance for different initial water saturations with linear $k_{r}$}
\label{fig:compare_2D_L}
\end{figure}

Now, we consider the scenario with both viscous and gravitational forces. The iteration performance for different initial water saturations with quadratic $k_{r}$ curves is shown in \textbf{Fig. \ref{fig:compare_2D_Q}}. For $\Delta t = 20.0 $ days, the maximum $\textrm{CFL}$ number is $\textrm{CFL} \approx $ 150. The water saturation profile at the end of simulation (300 days) for the initial water saturation 0.4 is plotted in \textbf{Fig. \ref{fig:compare_2D_Sw}}. Due to the non-equilibrium initial condition, gravity segregation is taking place in all cells at the beginning of the simulation. We also test the case with linear $k_{r}$ curves. The nonlinear iteration performance for different initial water saturations is shown in \textbf{Fig. \ref{fig:compare_2D_L}}. It can be seen that counter-current flow exacerbates the nonlinear convergence difficulty. The DBC method can significantly improve the performance, especially for the case with a non-equilibrium initial condition.

\begin{figure}[!htb]
\centering
\includegraphics[scale=0.6]{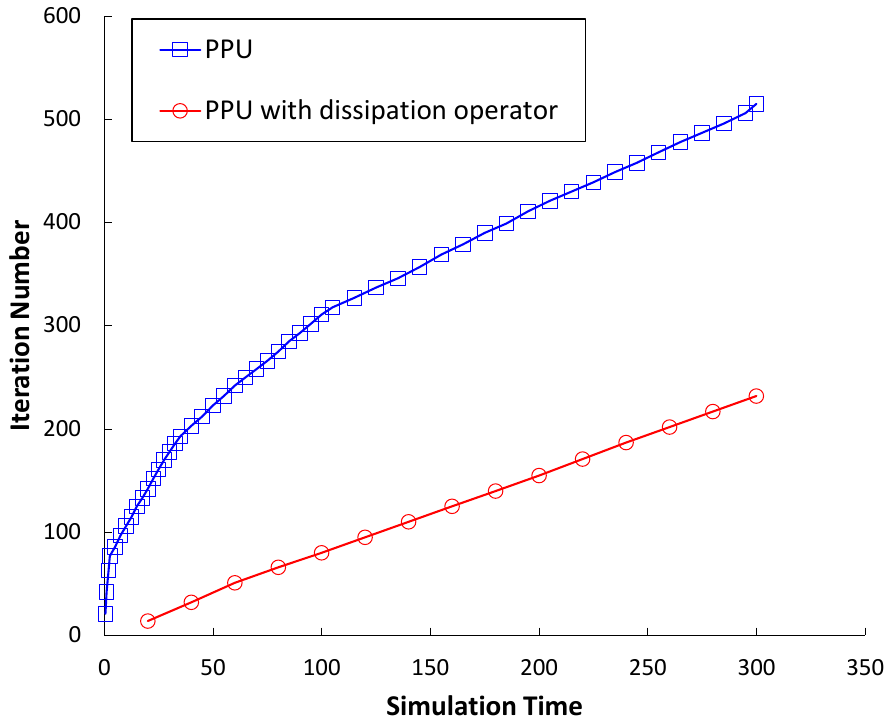}
\caption{Cumulative number of Newton iterations versus simulation time}
\label{fig:compare_2D_W2}
\end{figure}

We run a case with the injection rate changed to 1.0 $\textrm{m}^3/\textrm{D}$. The initial water saturation is 0.4, and the other parameters specified in the base model remain unchanged. The cumulative number of Newton iterations versus simulation time is presented in \textbf{Fig. \ref{fig:compare_2D_W2}}. Compared to the poor performance of the standard Newton method, the DBC method does not require any timestep cuts.

In this work the linear subproblems are solved using the Intel MKL PARDISO solver. Here we only study the impact of the DBC method on the nonlinear iterations. The dissipation operator may exhibit both detrimental and beneficial effects on performance of iterative linear solvers. On one hand, the addition of dissipation will increase the condition number of linear system. On the other hand, the dissipation operator can lead to more diagonal dominance and definiteness, improving the robustness of preconditioners. The investigation for the impacts of dissipation on efficiency of iterative linear solvers is beyond the scope of this paper and is subject to a future work.

\section{Summary}

The Newton method often fails for highly nonlinear problems with poor initial guesses. We develop a dissipation-based continuation (DBC) method for nonlinear two-phase flow and transport in porous media with combined viscous, gravitational, and capillary forces. The homotopy is constructed by adding numerical dissipation to the discrete flow equations, whereby a continuation parameter is used to gradually remove the dissipation. The DBC method acts as a globalization stage to obtain better initial guesses for the Newton process. Detailed analysis of single-cell and 1D problems helps to explain why the dissipation operator improves the convergence of nonlinear transport problems. We highlight a specific type of nonlinear convergence difficulty caused by low wave speeds around the saturation front. The typical behavior appears as a stagnation stage in the convergence history before the residual norm starts to decrease. An adaptive and local (cell based) strategy to determine the `optimum' dissipation coefficient is proposed. 

We demonstrate the effectiveness of the new nonlinear solver using several examples, including 1D scalar transport and 2D heterogeneous problems with fully-coupled flow and transport. The new solver exhibits superior convergence properties compared with the standard Newton solver used in reservoir simulation. The results show that the DBC method reduces the iteration count in all the cases investigated.
%
The artificial dissipation can help with spreading mass in the affected domain over the timestep and thus lead to faster nonlinear convergence.

The nonlinear DBC solver is flexible and can be integrated with other globalization techniques, such as trust-region and line-search algorithms, to guide the Newton direction and the step-length selection. In addition, the new solver can be readily applied to more complex multi-physics problems.

\section*{Acknowledgements}

This work was supported by the Stanford University Petroleum Research Institute for Reservoir Simulation (SUPRI-B). The authors thank Dr.\ Bradley Mallison for insightful comments on a draft of this manuscript.


\section*{Appendix A}

We follow the analysis presented in the work of Wang (2012).

\subsection*{Linear relative permeability curves}

Consider a 1D horizontal model and assume that linear relative permeability curves are used, i.e. 
\begin{equation} 
k_{rw} = S, \qquad k_{ro} = 1- S.
\end{equation}
If only viscous force is present, the water-phase flux can be written as
\begin{equation} 
f = \frac{MS}{1+\left ( M-1 \right )S}
\end{equation}
where $M$ is the viscosity ratio. In this simple setting, the residual equation for cell $i$ in discretized form is
\begin{equation} 
R = S_{i}^{n+1} - S_{i}^{n} + c\left ( f_{i+1/2}^{n+1} - f_{i-1/2}^{n+1} \right ),
\end{equation}
where c is a dimensionless timestep
\begin{equation} 
c = \frac{u_T \Delta t}{\Delta x}.
\end{equation}
The fluxes at the right and left interfaces are
\begin{equation}
f_{i+1/2}^{n+1} = \frac{M S_{i}^{n+1}}{1 + \left ( M-1 \right ) S_{i}^{n+1}},
\end{equation}
and
\begin{equation}
f_{i-1/2}^{n+1} = \frac{M S_{i-1}^{n+1}}{1 + \left ( M-1 \right ) S_{i-1}^{n+1}}.
\end{equation}

Hence, the residual can be expressed as
\begin{equation} 
\begin{split}
R &= \left ( S_{i}^{n+1} - S_{i}^{n}  \right ) + c\left ( f_{i+1/2}^{n+1} - f_{i-1/2}^{n+1} \right )  \\ 
 &= \left ( S_{i}^{n+1} - S_{i}^{n}  \right ) + cM\left ( \frac{S_{i}^{n+1}}{1+\left ( M-1 \right )S_{i}^{n+1}} - \frac{S_{i-1}^{n+1}}{1+\left ( M-1 \right )S_{i-1}^{n+1}} \right ) \\ 
 &= \left ( S_{i}^{n+1} - S_{i}^{n}  \right )
+ cM\frac{S_{i}^{n+1} - S_{i-1}^{n+1}}{\left ( 1+\left ( M-1 \right )S_{i}^{n+1} \right )\left ( 1+\left ( M-1 \right )S_{i-1}^{n+1} \right )}
\end{split}
\end{equation}
We obtain
\begin{equation} 
\frac{\partial R}{\partial S_{i}^{n+1}} = 1 + \frac{cM}{\left ( 1+\left ( M-1 \right )S_{i}^{n+1} \right )^2},
\end{equation}
and
\begin{equation} 
\frac{\partial R}{\partial S_{i-1}^{n+1}} = \frac{-cM}{\left ( 1+\left ( M-1 \right )S_{i}^{n+1} \right )\left ( 1+\left ( M-1 \right )S_{i-1}^{n+1} \right )}.
\end{equation}
If we take the initial condition $S^n = 0.0$ as initial guess for solution at $(n+1)$, then
\begin{equation} 
\frac{\partial R}{\partial S_{i}^{n+1}} = 1+ cM,
\end{equation}
and
\begin{equation} 
\frac{\partial R}{\partial S_{i-1}^{n+1}} = -cM.
\end{equation}
With the Newton method, the Jacobian for the first iteration is 
\begin{equation} 
\label{eq:jaco_1}
\begin{bmatrix}
1+cM &  &  & \\ 
-cM & 1+cM &  & \\ 
 & -cM & 1+cM & \\ 
 &  & \ddots  & \ddots 
\end{bmatrix}.
\end{equation}
Hence the timestep size $c$ is reflected in the wave speed. Assuming that the left-boundary condition is $S=S_0$, the residual (right hand side) for the first iteration is
\begin{equation} 
\begin{bmatrix}
-cM \frac{S_0}{1+\left ( M-1 \right )S_0} \\ 
0 \\ 
\vdots \\ 
0
\end{bmatrix}.
\end{equation}
Other boundary conditions result in different value for the residual of the left-most cell, whereas the residual terms of other cells will remain zero at the first iteration, regardless of the boundary condition. Thus, starting from the initial guess, $S=0$, the non-zero term in the residual is propagated downstream and the saturation is dispersive throughout the domain, since there is a non-zero element in every row of the lower off-diagonal part of the Jacobian (\ref{eq:jaco_1}).

When $M = 1$, the Jacobian has the following form
\begin{equation} 
\begin{bmatrix}
1+c &  &  & \\ 
-c & 1+c &  & \\ 
 & -c & 1+c & \\ 
 &  & \ddots  & \ddots 
\end{bmatrix}.
\end{equation}
In fact, for $M = 1$, the residual equation, $R$, becomes a linear function of the solution
\begin{equation} 
R = S_{i}^{n+1} - S_{i}^{n} + c\left ( S_{i}^{n+1} - S_{i-1}^{n+1} \right )
\end{equation}
and for this case, the Newton method will converge in one iteration regardless of the timestep size.

\subsection*{More general relative permeability curves}

Assume that the relative permeability curves are given by
\begin{equation} 
k_{rw} = S^{\alpha}, \qquad k_{ro} = \left (1-S \right )^{\beta}.
\end{equation}
where $\alpha \geq 1$ and $\beta \geq 1$. Then, the flux function can be written as
\begin{equation} 
f = \frac{S^{\alpha}/\mu_w}{S^{\alpha}/\mu_w + \left (1-S  \right )^{\beta }/\mu_o}.
\end{equation}
Hence, the derivative is
\begin{equation} 
\begin{split}
\frac{df}{dS} &= \frac{\alpha \frac{S^{\alpha}}{\mu_w}\left ( \frac{S^{\alpha}}{\mu_w} + \frac{\left ( 1-S \right )^{\beta}}{\mu_o}\right ) - \frac{S^{\alpha}}{\mu_w}\left ( \alpha \frac{S^{\alpha - 1}}{\mu_w} - \beta \frac{\left ( 1-S \right )^{\beta -1}}{\mu_o}\right )}{\left ( \frac{S^{\alpha}}{\mu_w} + \frac{\left ( 1-S \right )^{\beta}}{\mu_o} \right )^2} \\ 
 &= \frac{\alpha S^{\alpha -1}\left ( 1-S \right )^{\beta} + \beta S^{\alpha }\left ( 1-S \right )^{\beta -1}}{M \left ( S^{\alpha} + \frac{\left ( 1-S \right )^{\beta}}{M} \right )^2}.
\end{split}
\end{equation}
For $\alpha >1$, $\frac{df}{dS} = 0 $ for $S = 0.0$.

The residual is expressed as
\begin{equation} 
R = S_{i}^{n+1} - S_{i}^{n} + c \left ( f_{i+1/2}^{n+1} - f_{i-1/2}^{n+1} \right )
\end{equation}
Hence
\begin{equation} 
\frac{\partial R}{\partial S_{i}^{n+1}} = 1 + c \frac{df_{i+1/2}^{n+1}}{dS_{i}^{n+1}},
\end{equation}
and
\begin{equation} 
\frac{\partial R}{\partial S_{i-1}^{n+1}} = -c \frac{df_{i-1/2}^{n+1}}{dS_{i-1}^{n+1}}.
\end{equation}
Assume $\alpha >1$. When $S^n = 0$ everywhere and is taken as the initial guess for the next timestep $(n+1)$, then $\frac{\partial R}{\partial S_{i}^{n+1}} = 1$ and $\frac{\partial R}{\partial S_{i-1}^{n+1}} = 0$. So, the Jacobian matrix for the first Newton iteration is the identity matrix
\begin{equation} 
\begin{bmatrix}
1 &  &  & \\ 
0 & 1 &  & \\ 
 & 0 & 1 & \\ 
 &  & \ddots  & \ddots 
\end{bmatrix}.
\end{equation}
The timestep size $c$ does not appear in the first Jacobian matrix. Assuming that $f=1$ at the left boundary, the corresponding residual for the first iteration is
\begin{equation} 
\begin{bmatrix}
-c\\ 
0\\ 
\vdots \\ 
0
\end{bmatrix}.
\end{equation}
Since the Jacobian matrix is the identity, we can see that the non-zero term in the residual cannot propagate more than one cell after the first iteration. On the other hand, even though $S^n = 0.0$ is the initial condition, if we take the initial guess of the solution for the current timestep, $S^{n+1,0}$, as $S^{\ast} \neq 0.0$, we then have
\begin{equation} 
\frac{\partial R}{\partial S_{i}^{n+1}} = 1 + c {f}'^{\ast},
\end{equation}
and
\begin{equation} 
\frac{\partial R}{\partial S_{i-1}^{n+1}} = -c {f}'^{\ast},
\end{equation}
Hence, the Jacobian matrix for the first iteration is 
\begin{equation} 
\label{eq:jaco_2}
\begin{bmatrix}
1+c {f}'^{\ast} &  &  & \\ 
-c {f}'^{\ast} & 1+c {f}'^{\ast} &  & \\ 
 & -c {f}'^{\ast} & 1+c {f}'^{\ast} & \\ 
 &  & \ddots  & \ddots 
\end{bmatrix}.
\end{equation}
where ${f}'^{\ast}$ is $\frac{df}{dS}$ evaluated at $S^\ast $. The timestep size $c$ appears in the Jacobian matrix if ${f}'^{\ast} \neq 0$. Assuming that the left boundary condition is $f=1$, the corresponding residual vector for the first iteration is
\begin{equation} 
\begin{bmatrix}
S^\ast + c\left ( f^\ast -1 \right )\\ 
S^\ast \\ 
\vdots \\ 
S^\ast
\end{bmatrix}
\end{equation}
the solution update, $\delta S$, is obtained by solving the linear system $J \delta S = - R$, and then $S^\ast + \delta S$ serves as the starting point for the second Newton iteration.

Now we show that for the first Newton iteration, if ${f}'^{\ast} > 0 $, the resulting $S^\ast + \delta S$ decreases monotonically as $i$ increases from 1 to $N$, where $N$ is the number of cells. Since the Jacobian matrix (\ref{eq:jaco_2}) is lower triangular, we can solve the elements in $\delta S = \left [ \delta S_1, \delta S_2, \cdots, \delta S_N \right ]^T$ one by one
\begin{equation} 
\begin{bmatrix}
1+c {f}'^{\ast} &  &  & \\ 
-c {f}'^{\ast} & 1+c {f}'^{\ast} &  & \\ 
 & -c {f}'^{\ast} & 1+c {f}'^{\ast} & \\ 
 &  & \ddots  & \ddots 
\end{bmatrix}
\begin{bmatrix}
\delta S_1\\ 
\delta S_2 \\ 
\vdots \\ 
\delta S_N
\end{bmatrix}
= -\begin{bmatrix}
S^\ast + c\left ( f^\ast -1 \right )\\ 
S^\ast \\ 
\vdots \\ 
S^\ast
\end{bmatrix}
\end{equation}
First, we obtain the solution update for the first grid-block
\begin{equation} 
\delta S_1 = - \frac{S^\ast + c\left ( f^\ast -1 \right )}{1+c{f}'^{\ast}}.
\end{equation}
Hence
\begin{equation} 
S^\ast + \delta S_1 = \frac{c\left ( S^\ast {f}'^{\ast}- f^\ast + 1 \right ) }{1+c{f}'^{\ast}} >0 
\end{equation}
For $i \geq 2$
\begin{equation} 
-c{f}'^{\ast}\delta S_{i-1} + \left ( 1+c{f}'^{\ast} \right )\delta S_i = - S^\ast 
\end{equation}
Therefore
\begin{equation}
\label{eq:proof_1} 
\delta S_i + S^\ast = \frac{c{f}'^{\ast}}{1+c{f}'^{\ast}}\left ( \delta S_{i-1} + S^\ast \right ),
\end{equation}
and since ${f}'^{\ast} > 0$,
\begin{equation} 
0<\frac{c{f}'^{\ast}}{1+ c{f}'^{\ast}}<1.
\end{equation}
It follows that
\begin{equation} 
\delta S_i + S^\ast < \delta S_{i-1} + S^\ast 
\end{equation}
Therefore, $S^\ast + \delta S$ decreases monotonically as $i$ increases.

The solution can be written as
\begin{equation} 
\delta S_i + S^\ast = \left ( \frac{c{f}'^{\ast}}{1+ c{f}'^{\ast}} \right )^{i-1} \left ( \delta S_1 + S^* \right ),
\end{equation}
and since $S^\ast + \delta S_1 > 0 $, then
\begin{equation} 
S^\ast + \delta S_i > 0   \qquad \forall i\geq 2.
\end{equation}
It is shown that the saturation solution from the first iteration is positive (and less than unity) in every cell of the 1D domain. From Eq. (\ref{eq:proof_1}), we can see that for any initial guess, $S^\ast $, if ${f}'^{\ast} > 0 $, the first Newton iteration yields a saturation distribution that is monotonic and positive.

\section*{Appendix B}

The Newton-Mysovskikh theorem (Ortega and Rheinboldt 1970) states that the Newton method is guaranteed to converge if the convergence ratio $\left | R(S){R}''(S) \right |/ \left | {R}'(S) \right |^{2} < 2$ is maintained locally, assuming existence of a solution. The residual $R(S)$ in Eq. (\ref{Eq:Resi}) is taken to be a $C^2$ function of the solution $S$. A convergence ratio satisfying the criterion indicates that it is now in the contraction region around the root, and the iterations will converge (Wang and Tchelepi 2013).

Consider a single cell problem with immiscible two-phase transport. The saturations, $S_L$ and $S_R$, are the left and right boundary conditions. The residual form of the conservation law can be written as
\begin{equation} 
\label{Eq:Resi}
R(S^{n+1}) = \left ( S^{n+1} - S^{n} \right ) + \frac{\Delta t}{\Delta x}\left ( F_R^{n+1} - F_L^{n+1} \right ) 
\end{equation}
where the initial condition is $S^n=0$. The derivatives of the residual are
\begin{equation} 
{R}' = \frac{dR}{dS^{n+1}} = 1 + \frac{\Delta t}{\Delta x} \left ( \frac{\partial F_R(S^{n+1}; S_R)}{\partial S^{n+1}} - \frac{\partial F_L(S_L; S^{n+1})}{\partial S^{n+1}} \right )
\end{equation}
\begin{equation} 
{R}'' = \frac{d^2R}{d(S^{n+1})^2} = \frac{\Delta t}{\Delta x} \left ( \frac{\partial^2 F_R(S^{n+1}; S_R)}{\partial (S^{n+1})^2} - \frac{\partial^2 F_L(S_L; S^{n+1})}{\partial (S^{n+1})^2} \right )
\end{equation}

\begin{figure}[!htb]
\centering
\subfloat[$C_g= -0.01 $ and $\beta = 2 $]{
\includegraphics[scale=0.7]{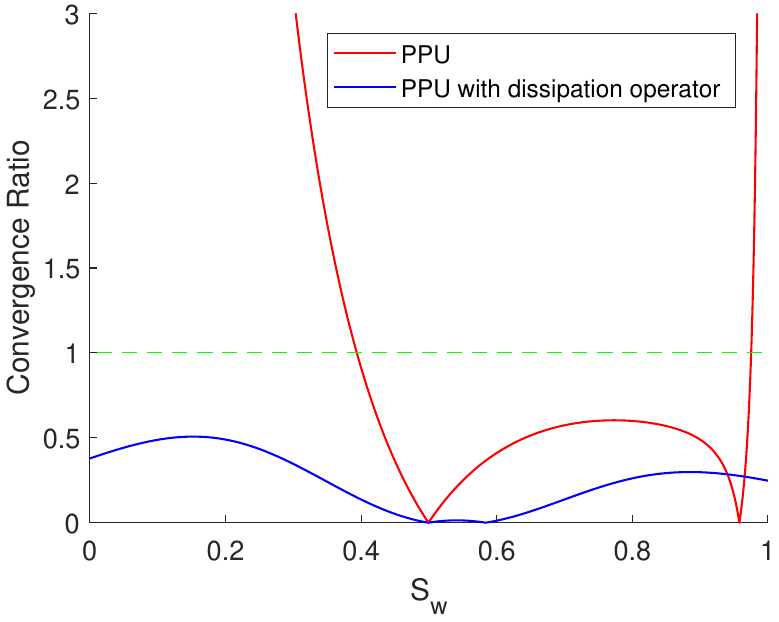}}
\\
\subfloat[$C_g=-5$ and $\beta = 4 $]{
\includegraphics[scale=0.7]{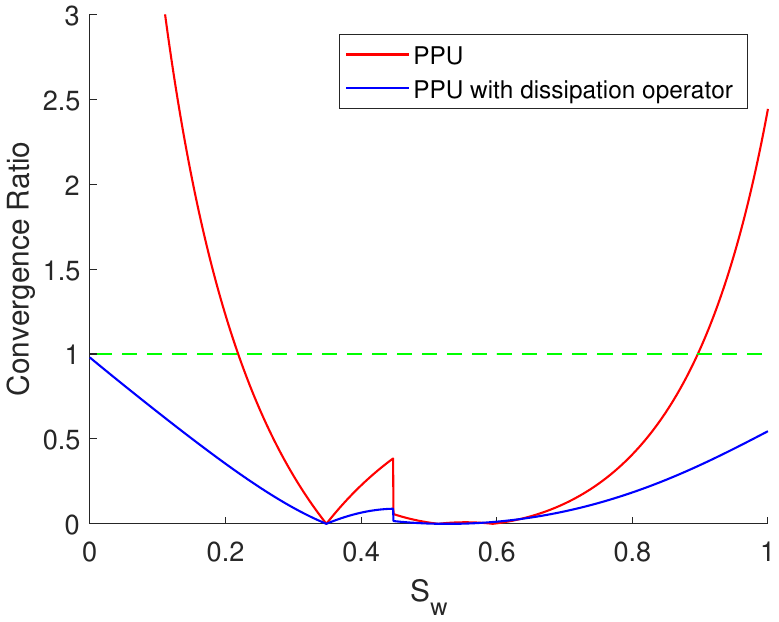}}
\caption{Convergence ratios of PPU with and without the dissipation operator.}
\label{fig:CR_VG}
\end{figure}

We compare the convergence ratios of the PPU numerical flux with and without the dissipation operator. We take $u_T=1$, $M=1$ and the ratio $\frac{\Delta t}{\Delta x}$ is 500. The boundary conditions are set to $S_L=1.0$ and $S_R=0.0$. The convergence ratios versus the cell saturation for the cases with different $C_g$ and $\kappa = 1 $ are plotted in \textbf{Fig. \ref{fig:CR_VG}}. From the figure of the gravitational-dominated case, we observe that the convergence ratio of PPU is quite different between the two sides of the unit-flux point due to the counter-current flow. It can be seen that the dissipation term exhibits a noticeable regularizing effect on the residuals of the flow equation. The convergence ratio for PPU with the dissipation operator is much smaller than the standard scheme. This demonstrates that the dissipation operator leads to a favorable property for nonlinear convergence. 

\section*{References}

Aziz, K., Settari, A., 1979. Petroleum Reservoir Simulation. Chapman \& Hall.

Allgower, E.L. and Georg, K., 1993. Continuation and path following. Acta numerica, 2, pp.1-64.

Brenier, Y. and Jaffré, J., 1991. Upstream differencing for multiphase flow in reservoir simulation. SIAM journal on numerical analysis, 28(3), pp.685-696.

Brune, P.R., Knepley, M.G., Smith, B.F. and Tu, X., 2015. Composing scalable nonlinear algebraic solvers. SIAM Review, 57(4), pp.535-565.

Brown, D.A. and Zingg, D.W., 2016. A monolithic homotopy continuation algorithm with application to computational fluid dynamics. Journal of Computational Physics, 321, pp.55-75.

Brown, D.A. and Zingg, D.W., 2017. Design and evaluation of homotopies for efficient and robust continuation. Applied Numerical Mathematics, 118, pp.150-181.

Cai, X.C., Keyes, D.E. and Marcinkowski, L., 2002. Non‐linear additive Schwarz preconditioners and application in computational fluid dynamics. International journal for numerical methods in fluids, 40(12), pp.1463-1470.

Chen, Z., Huan, G. and Ma, Y., 2006. Computational methods for multiphase flows in porous media (Vol. 2). Siam.

Cogswell, D.A. and Szulczewski, M.L., 2017. Simulation of incompressible two-phase flow in porous media with large timesteps. Journal of Computational Physics.

Deuflhard, P., 2004. Newton methods for nonlinear problems: affine invariance and adaptive algorithms (Vol. 35). Springer Science $\&$ Business Media.

ECLIPSE Technical Description 2008. Houston: Schlumberger GeoQuest.

Jenny, P., Tchelepi, H.A. and Lee, S.H., 2009. Unconditionally convergent nonlinear solver for hyperbolic conservation laws with S-shaped flux functions. Journal of Computational Physics, 228(20), pp.7497-7512.

Keller, H.B., 1977. Numerical Solution of Bifurcation and Nonlinear Eigenvalue Problems. Applications of bifurcation theory, pp.359-384.

Knoll, D.A. and Keyes, D.E., 2004. Jacobian-free Newton–Krylov methods: a survey of approaches and applications. Journal of Computational Physics, 193(2), pp.357-397.

Kwok, F. and Tchelepi, H.A., 2008. Convergence of implicit monotone schemes with applications in multiphase flow in porous media. SIAM Journal on Numerical Analysis, 46(5), pp.2662-2687.

Li, B. and Tchelepi, H.A., 2015. Nonlinear analysis of multiphase transport in porous media in the presence of viscous, buoyancy, and capillary forces. Journal of Computational Physics, 297, pp.104-131.

Lee, S.H., Efendiev, Y. and Tchelepi, H.A., 2015. Hybrid upwind discretization of nonlinear two-phase flow with gravity. Advances in Water Resources, 82, pp.27-38.

M{\o}yner, O., 2016, August. Nonlinear solver for three-phase transport problems based on approximate trust regions. In ECMOR XV-15th European Conference on the Mathematics of Oil Recovery.

Ortega, J.M. and Rheinboldt, W.C., 1970. Iterative solution of nonlinear equations in several variables (Vol. 30). Siam.

Pulliam, T., 1986. Artificial dissipation models for the Euler equations. AIAA journal, 24(12), pp.1931-1940.

Forsyth Jr, P.A. and Sammon, P.H., 1986. Practical considerations for adaptive implicit methods in reservoir simulation. Journal of Computational Physics, 62(2), pp.265-281.

Peaceman, D.W., 2000. Fundamentals of numerical reservoir simulation (Vol. 6). Elsevier.

Sommese, A.J. and Wampler, C.W., 2005. The Numerical solution of systems of polynomials arising in engineering and science. World Scientific.

Skogestad, J.O., Keilegavlen, E. and Nordbotten, J.M., 2013. Domain decomposition strategies for nonlinear flow problems in porous media. Journal of Computational Physics, 234, pp.439-451.

Watson, L.T., 1990. Globally convergent homotopy algorithms for nonlinear systems of equations. Nonlinear Dynamics, 1(2), pp.143-191.

Wang, X., 2012. Trust-Region Newton Solver for Multiphase Flow and Transport in Porous Media (Doctoral dissertation, PhD Thesis, Stanford University).

Wang, X. and Tchelepi, H.A., 2013. Trust-region based solver for nonlinear transport in heterogeneous porous media. Journal of Computational Physics, 253, pp.114-137.

Younis, R., Tchelepi, H.A. and Aziz, K., 2010. Adaptively Localized Continuation-Newton Method--Nonlinear Solvers That Converge All the Time. SPE Journal, 15(02), pp.526-544.

\end{document}